\documentclass[12pt]{article} 
\usepackage{latexsym} 
\usepackage{amssymb} 
\usepackage{amsmath,amsfonts} 
\usepackage{graphicx} 
\usepackage{braket} 
\usepackage{color}

\catcode`\@=11 
\def\numberbysection{\@addtoreset{equation}{section} 
        \def\theequation{\thesection.\arabic{equation}}} 
 
\def\be{\begin{equation}} 
\def\ee{\end{equation}} 
\def\ba{\begin{eqnarray}} 
\def\ea{\end{eqnarray}} 
 
\def\ov{\overline}

\def\Z{\mathbb{Z}}

\def\nl{\nonumber \\}

\def\de{\partial} 
\def\wt{\widetilde} 
\def\wh{\widehat} 
 
\def\dag{\dagger} 
 
\def\a{\alpha} 
\def\b{\beta}

\def\D{\Delta}

\def\eps{\varepsilon}

\def\l{\lambda} 
\def\L{\Lambda}

\def\p{\pi} 
 
\def\r{\rho} 
\def\s{\sigma} 
\def\S{\Sigma}

\def\vf{\varphi} 
\def\F{\Phi} 
\def\c{\chi} 
 
\def\W{\Omega} 
 
\def\th{\theta}

\def\u1{\widehat{U(1)}} 
\def\su2{\widehat{SU(2)}_1}

\numberbysection 
 
\begin{document} 
 
\begin{titlepage} 
\begin{center} 

\vskip .6 in 
{\LARGE Stability of Topological Insulators} 
 
{\LARGE with Non-Abelian Edge Excitations} 
 
\vskip 0.2in 
Andrea CAPPELLI ${}^{(a)}$\\ 
 
Enrico RANDELLINI ${}^{(a,b)}$\\ 
 
{\em ${}^{(a)}$ INFN, Sezione di Firenze}\\ 
{\em ${}^{(b)}$ Dipartimento di Fisica, Universit\`a di Firenze\\ 
Via G. Sansone 1, 50019 Sesto Fiorentino - Firenze, Italy} 
\end{center} 
\vskip .2 in 
\begin{abstract} 
Chiral-antichiral pairs of non-Abelian Hall states, like the
Pfaffian, Read-Rezayi and NASS states, can be used to model
two-dimensional time-reversal invariant topological insulators. Their
stability was shown to be associated to the presence of a $\Z_2$
anomaly and characterized by the same $\Z_2$ index introduced for free
fermion and Abelian systems.  In this work, we continue the stability
analysis by providing the form of time-reversal invariant interactions
that gap the non-Abelian edge excitations. Our approach is based
on the description of non-Abelian states as projections of
corresponding ``parent'' Abelian states.

\end{abstract} 
 
%\pacs{73.43.Cd, 11.25.Hf, 73.23.Hk, 73.43.Jn} 
\vfill 
 
\end{titlepage} 
\pagenumbering{arabic}

%-1--------------------------------------------- 
 
\section{Introduction} 
 
Topological states of matter in two and three dimensions are currently
investigated both theoretically \cite{qz-rev} and experimentally
\cite{molen} \cite{3d-ti}, and new systems have been suggested in
combined two-three dimensional geometries \cite{3d-interact}.  Of
particular interest are the time-reversal invariant systems, such as
the topological insulators, that can be realized in absence of
external magnetic fields.  They are characterized by gapful
excitations in the bulk and massless charged edge excitations.
 
Two-dimensional topological insulators with interacting fermions,
i.e. different from band insulators, can be modeled by pairs of
quantum Hall states carrying opposite spin and chirality, such that
time-reversal invariant systems are obtained \cite{ls}.  Assuming the
existence of a suitable bulk Hamiltonian that gives rise to both sets
of edge excitations, the question remains of the interactions between
electrons of opposite chirality at the edge, that can gap completely
the system and made it topologically trivial. Some interactions may be
forbidden by time-reversal symmetry, possibly combined with other
discrete symmetries, leading to a so-called symmetry protected
topological phase \cite{wen}. Thus, the stability analysis is
particularly relevant in this kind of theoretical modeling \cite{ls}
\cite{chamon} \cite{vish}.

In our earlier paper \cite{cr}, a stability criterium based on
symmetry arguments \cite{ls} was generalized to
time-reversal invariant topological insulators with non-Abelian edge
excitations (see also \cite{ringel}).  In the present work, we
complement this analysis with the study of the allowed interactions.
Analogous analyses are currently developed for interacting
two-dimensional topological superconductors \cite{ryu} and for
two-dimensional systems on the surface of three-dimensional
topological insulators \cite{3d-interact} \cite{levin} \cite{vish2}.

\subsection{Flux insertion argument and edge interactions}

\begin{figure}[t] 
\begin{center} 
\includegraphics[width=10cm]{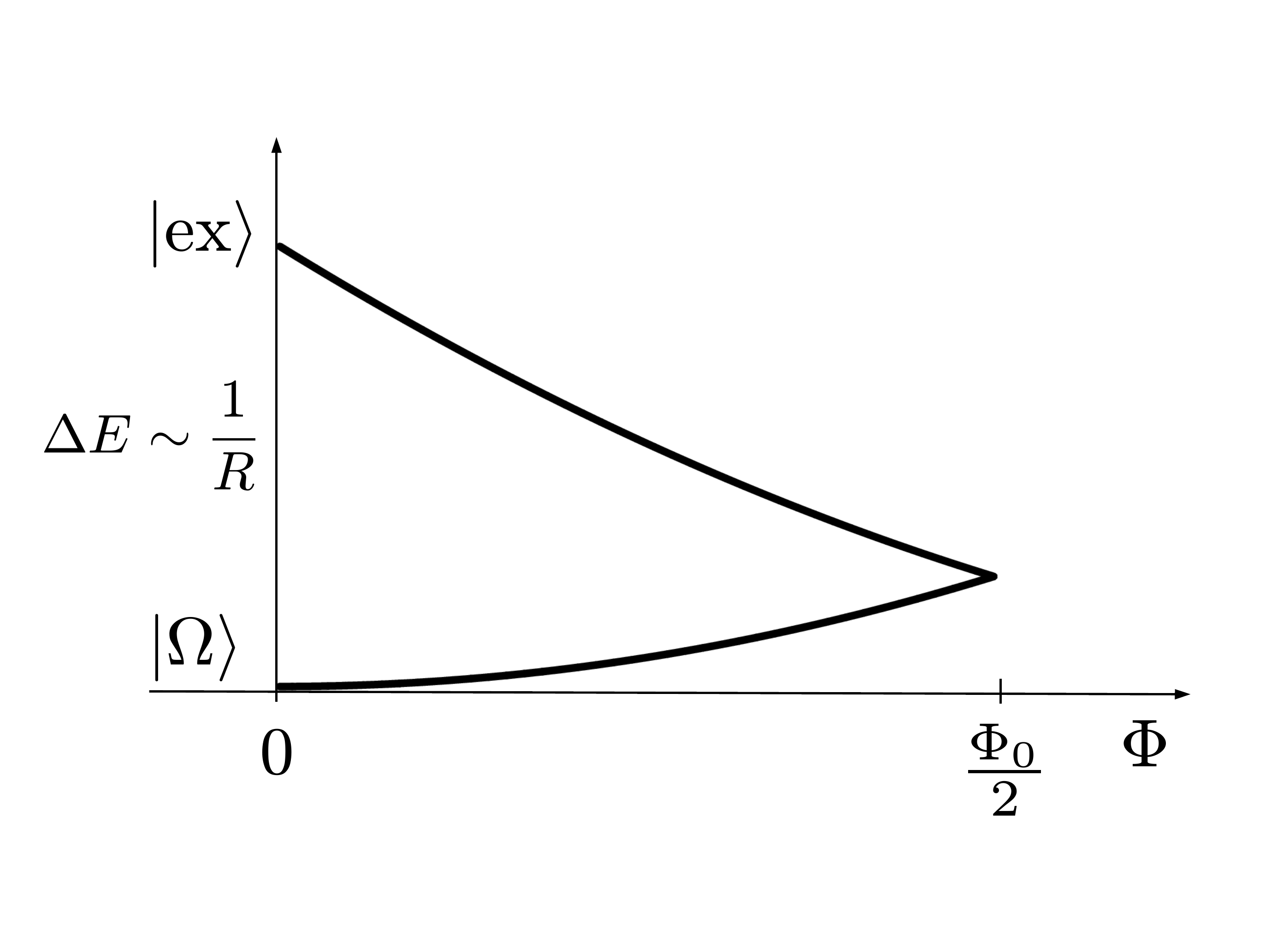} 
\caption{Kramers degeneracy at half flux quantum.} 
\label{fig1} 
\end{center} 
\end{figure}

The stability analysis of topological insulators was based on two
approaches: the first used symmetry and topology arguments and the
Kramers theorem; the second involved a systematic study of edge
interactions that do not break time-reversal symmetry.  In the first
approach, pioneered by Fu, Kane and Mele \cite{kane} \cite{tbt}
\cite{qz}, one considers the response of the system, say in the
annulus geometry, to the insertion of flux quanta at the center -- a
refinement of the well-known Laughlin argument for the quantized Hall
current \cite{laugh} \cite{cdtz}.  For example, in the case of a
single spinful free fermion mode, the addition of half flux $\Phi_0/2$ let the
ground state $\vert \Omega\rangle$ 
evolve into a spin $S=1/2$ neutral edge excitation,
corresponding to a change of the $\Z_2$ index $(-1)^{2S}$:

\be \Phi=0:
\ (-1)^{2S} =1 \ \ \longrightarrow
\ \ \Phi=\frac{\Phi_0}{2}:\ (-1)^{2S}=-1.
\label{change s-p} 
\ee 
In time-reversal invariant systems, the Kramers theorem
implies that this edge state is degenerate with another $S=1/2$
state $\vert {\rm ex}\rangle$
coming from the excitations (see Fig.\ref{fig1}).  This
degeneracy is robust with respect to any perturbation that preserves
time-reversal symmetry, and it implies that  ground state and
excitations at $\Phi=0$ are separated by a gap $\D E=O(1/R)$, that
vanishes in the large volume limit. Therefore, the time-reversal
invariant topological insulator of one free fermion has
gapless edge excitations and is topologically non trivial.
The argument clearly extends to $N$ free fermion systems that are
stable (resp. unstable) for $N$ odd (even).

The flux insertion argument was generalized by Levin and Stern
\cite{ls} to systems with Abelian edge excitations, that can be described
by multicomponent Luttinger liquids \cite{wen-book}. The $\Z_2$ index
was show to extend as follows: 
\be 
(-1)^{2\Delta S}, \qquad 2\Delta
S=\frac{\nu^\uparrow}{e^*},
\label{ls-index} 
\ee 
where $\nu^\uparrow$ is the Hall filling fraction and $e^*$ is the
minimal fractional charge, in units of $e$, of one chiral component 
(equivalently, the dimensionless ratio of the spin Hall conductivity
and minimal charge $\sigma_{sH}/e^*$).  In a system with anyonic
excitations, the ratio (\ref{ls-index}) correctly measures the spin $\D S$ of
the smallest electron excitation created at the edge by adding half
fluxes. An odd (resp. even) ratio corresponds to stable (unstable) systems,
generalizing the number of fermion modes in the non-interacting
case.
 
The second stability analysis was based on studying the possible
time-reversal invariant electron interactions at the edge \cite{ls}
\cite{chamon}. Their general expressions in multi-component Abelian
systems can be written using vertex operators of the bosonic
conformal field theory in the so-called $K$-matrix formalism.  It was
found in Ref. \cite{ls} that the stability of the edge modes is again
determined by the index (\ref{ls-index}): when this is positive,
i.e. the Kramers degeneracy argument does not hold, there are enough
interactions for gapping all edge modes; conversely, if the index is
negative, i.e. there is Kramers degeneracy, one mode remains gapless.
In conclusion, both approaches of studying flux insertions and
interactions led to the same conclusions about the stability of
topological insulators with Abelian edges.

\subsection{Non-Abelian topological insulators}

In a recent paper \cite{cr}, we were able to extend the first kind of
stability analysis to topological insulators involving pairs of
non-Abelian edge excitations, like the Pfaffian \cite{mr}, the
Read-Rezayi \cite{rr} and the non-Abelian Spin Singlet States
\cite{nass}.  We first obtained the grand-canonical partition function
of the conformal field theory describing chiral-antichiral pairs of
edge excitations, by dwelling on earlier results \cite{cz} \cite{cv}.
Then, we used the partition function for discussing the flux argument
in full generality for any interacting system.

Quantum Hall edge states possess neutral modes that can be Abelian or 
non-Abelian. The corresponding conformal theories have the 
affine symmetry $U(1) \times G/H$, where 
$U(1)$ is the charge symmetry and $G$ is another (non-Abelian) symmetry 
characterizing the neutral part (possibly a coset $G/H$) \cite{cv}. 
The electron field is represented by the product 
of a chiral vertex operator $V=e^{i\a \varphi}$ for the charge part 
and the neutral field $\psi_e$ of the $G/H$ theory: 
\be 
\Psi_e=e^{i\a \varphi}\, \psi_e\ . 
\label{el-field} 
\ee 
Even in non-Abelian theories, the field $\psi_e$ should have Abelian
fusion rules with all fields in the theory, i.e. should be a so-called
simple current \cite{cft}. This field can be used to build the
partition function on the spacetime torus that is modular invariant,
i.e. symmetric under discrete coordinate changes that respect the
double periodicity \cite{cft}. It was found that the partition
function of any quantum Hall state is uniquely determined by two
inputs: the choice of neutral $G/H$ theory and of Abelian field
$\psi_e$ in the theory that represents the neutral part of the
electron \cite{cv}.

The study of transformations of the partition function under
 flux insertions \cite {cr} showed that the neutral sectors do not
play any role, and that the stability depends on a pair of numbers
$(k,p)$ parameterizing the charge spectrum, namely the value of the
minimal charge, $e^*=1/p$, and of the would-be chiral Hall conductivity
(spin conductivity), $\nu^\uparrow =k/p$.  The minimal spin $\D S$
excitation created by flux insertions was found to match the
Levin-Stern index (\ref{ls-index}), again
\be 
2\Delta S= \frac{\nu^\uparrow}{e^*}=k,  \qquad
(-1)^{2\Delta S}=(-1)^{k}.
\label{ls-index-k} 
\ee 
In particular, the simpler topological insulator made by a pair of
Pfaffian states is unstable, since it corresponds to
$e^*=1/4$ and $\nu^\uparrow=1/2$, i.e. to $(k,p)=(2,4)$.

In the same paper, we emphasized that the change of index under flux
insertions expresses a discrete $\Z_2$ anomaly, that is the
non-conservation of the (edge) spin parity $(-1)^{2S}$ of the
fermionic system \cite{kane} \cite{ringel} \cite{anom}.  Indeed, this
anomaly is the remnant of the $U(1)_S$ continuous anomaly of the spin
Hall effect after the inclusion of relativistic corrections, such as
spin-orbit interaction, that break the spin conservation explicitly
(but keep time-reversal invariance) \cite{qz-rev}.

The behaviour of the partition function under modular transformations
was also analyzed \cite{cr}. In presence of fermionic excitations,
this function always possesses four parts (spin sectors), called
the Neveu-Schwarz and Ramond sectors and their tildes; these are
characterized by antiperiodic and periodic boundary
conditions for the fermion field in each direction \cite{cft}. It was
found that the half-flux insertions and the modular transformations
$S$ and $T$ map the four spin sectors among themselves. In the case of
unstable non-anomalous systems, time-reversal symmetry allows to sum
the four sectors together, leading to the complete modular invariant
$Z_{\rm Ising}$; on the contrary, in the presence of $\Z_2$ anomaly
they cannot be summed up, leading to four independent partition
functions, as follows: \ba Z_{\rm Ising}&=&Z^{NS} \! + Z^{\wt{NS}}\!
+Z^R\! +Z^{\wt{R}}, \qquad\qquad\ \ k\ {\rm even,\ unstable}, \nl
Z_{\rm TR}&=&\left(Z^{NS},Z^{\wt{NS}}, Z^R,Z^{\wt{R}} \right),
\qquad\qquad\quad k\ {\rm odd,\ stable}.
\label{non-inv.pf} 
\ea 
Therefore, the $\Z_2$ spin parity anomaly was associated
to a discrete gravitational anomaly, i.e. to a lack of complete modular
invariance of the torus partition function \cite{rz}.  

%-1.3----------------------------------
\subsection{Interactions in non-Abelian topological insulators}

In our previous paper \cite{cr} we did not discuss the allowed edge
interactions, thus we could not check whether a non-anomalous system,
not protected by the Kramers theorem, does actually become fully
gapped. In this paper, we provide the expressions of a sufficient set
of gapping interactions passing all physics tests.
  
We use the known result that some non-Abelian states can be described
as projections of corresponding ``parent'' Abelian states
\cite{cgt1}\cite{cgt2}. For example, the $(331)$ Halperin state of
distinguishable electrons is related to the Pfaffian state by the
projection onto states of identical electrons, e.g. by
antisymmetrizing over wavefunction coordinates.  Since this projection
does not affect the time-reversal symmetry of states and operators, it
maps time-reversal invariant interactions between the two theories.
Clearly, the value of the Levin-Stern index is equal in the
unprojected (Abelian) and projected (non-Abelian) theories \cite{cr}.
Through this mapping the earlier study of Abelian edge interactions
\cite{ls} \cite{chamon} \cite{vish} can be extended to non-Abelian
cases, where it provides the possible electrons interactions.  It
turns out that these are not sufficient to completely gap the system,
but further neutral quasiparticle interactions can be introduced
that do this task when the Levin-Stern index is one.
 
The plan of the paper is the following.  In Section two, we recall the
analysis of time-reversal interactions in multicomponent Abelian
theories.  In Section three, we introduce and use 
the projection from the Abelian $(331)$ state into the Pfaffian to
obtain the non-Abelian interactions. In analyzing their properties,
special attention is paid to normal ordering of fields and to the
absence of spontaneous breaking of time-reversal symmetry. In Section
four, we generalize this results to the $\Z_k$-parafermionic
Read-Rezayi states.  In Section five, we similarly study topological
insulators made by pairs of non-Abelian spin-singlet (NASS) states,
which turn out to be all unstable.  Finally, in Section six we present
our conclusions.

%-2-------------------------------

\section{Time-reversal invariant interactions in Abelian theories}
 
We start by recalling the conformal field theory description of
bosonic fields within the $K$-matrix formalism \cite{wen-book}. In the
case of topological insulators, the lattice of excitations is doubled
to account for the pairs of fields with opposite spin and chirality, as
described in \cite{ls} \cite{chamon} (we adopt the notation and
conventions of \cite{chamon}).  We introduce the $2N\times 2N$
symmetric invertible matrix $\mathcal{K}$ with integer components:
time-reversal symmetry determines its form to be,
\be
\label{Kmatrix} 
\mathcal{K}= \begin{pmatrix} K& W\\W^{T}&-K \end{pmatrix}, 
\ee 
where $K$ is the usual symmetric $N\times N$ matrix of Abelian Hall systems
and $W^T=-W$.

A generic (multi)-electron excitation is specified by a vector $\L$
with $2N$ integer components, such that its statistics, $\th/\pi=\L^T
\mathcal{K} \L$, and charge, $Q=\L^T\rho$, are integer-valued, where
$\r$ is the so-called charge vector. In our basis, 
this is made of two equal $N$-dimensional vectors, 
$\r=(\r^\uparrow,\r^\downarrow)$, $\r^\uparrow=\r^\downarrow=(1,\dots,1)$; 
then, the elementary electron excitations correspond to the basis
vectors $\L=e_i$, that are equal to one in the $i$-th position and zero
elsewhere, $i=1,\dots,2N$.  The electrons are represented by normal
ordered vertex operators of the $2N$-component bosonic field
$\F(t,x)$, as follows:
\be
\label{ele} 
\Psi_i^{\dag}(t,x)= : \exp \left(-i e_i^T \mathcal{K} \F(t,x)
\right):, \qquad i=1,\cdots,2N.  
\ee 
If $W=0$ in (\ref{Kmatrix}), the first $N$ operators, $i=1,\cdots,N$,
represent chiral spin-up electrons and the second $N$ ones antichiral
spin-down electrons; if $W\neq 0$, the first (resp. second) $N$ operators 
describe electrons with spin up (down) with mixed chiralities.

The time-reversal $\cal{T}$ transformations act on the bosonic field 
as follows \cite{ls}:
\be 
\label{trf} 
{\cal T}\ \F(t,x)\ {\cal T}^{-1}=\S_1\F(-t,x)+
\p\mathcal{K}^{-1}\S_{\downarrow}\r,
\ee 
where 
\be 
\label{sig1} 
\S_1=\begin{pmatrix} \mathbf{ 0} & \mathbf{ 1} \\ 
\mathbf{1} & \mathbf{0} 
         \end{pmatrix}, 
\ \ \ \
\S_{\downarrow}=\begin{pmatrix} \mathbf{ 0} & \mathbf{ 0} \\ 
 \mathbf{0} & \mathbf{1} 
\end{pmatrix}, 
\ee 
are $2N\times 2N$ block matrices. Time-reversal symmetry implies 
$\mathcal{K}= - \S_1\ \mathcal{K} \ \S_1,$ and $\r = \S_1 \r$. 
 
The time-reversal transformations of the basic fermionic fields (\ref{ele})
are as follows, keeping in mind that ${\cal T} $ is antiunitary,
\be
\label{tr-ep}
 {\cal T}: \  \  \   \Psi_{i}^{\dag}=\ 
: \exp \big(-i e_i^T \mathcal{K} \F  \big): 
\ \ \to \ \ : 
\exp \big(-i(\S_1 e_i)^T \mathcal{K} \F 
-i\p e_i^T\S_{\downarrow}\r\big) :,
\ee 
namely 
\be
\label{tr-ele} 
{\cal T}: \ \ \  \Psi_{i}^{\dag}\ \ \to \ \ \Psi_{i+N}^{\dag},\qquad \qquad \qquad
\Psi_{i+N}^{\dag}\ \ \to \ \ -\Psi_{i}^{\dag}, \qquad
i=1,\dots,N.
\ee

%-2.1----------------------------------------------

\subsection{Time-reversal symmetric interactions} 

The Hamiltonian $H_{\rm int}$ of electronic edge interactions 
is expressed in terms of vertex operators $U_{\L_i}$ as follows:
\ba
&&H_{\rm int}=\int dt\ \sum_i\, g_i\, U_{\L_i} + \text{h.c.},
\nl
&&U_{\L_i}(t,x)=: \exp \big(-i \L_i^T \mathcal{K} \F(t,x)\big):,
\label{H.int}
\ea 
where $\L_i$ are integer vectors subjected to the conditions
specified below. The coupling constant $g_i$ can be complex and space
dependent to account for interactions at impurities, possibly leading
to $g_i\to\infty$, such that both relevant and irrelevant interactions
should be considered.

The condition obeyed by the $\L_i$ for admissible interactions are 
\cite{ls} \cite{chamon}:

i) charge neutrality, 
\be
Q=\L_i^T \r=0 ;
\label{neutral}
\ee

ii) mutual locality of all interactions (Haldane null vector criterion 
\cite{hal}),
\be 
\label{Haldane} 
\frac{\th}{\pi}=\L_i ^T\mathcal{K} \L_j=0, \ \ \ \forall i,j,
\ee 
such that each interaction can effectively freeze one Abelian edge mode;

iii) time-reversal invariance of $H_{\rm int}$, 
\be
\S_1 \L_i =\pm \L_i, \qquad \L_i^T\ \S_{\downarrow}\ \r={\rm even}, 
\quad \forall i,
\label{trinv}
\ee
as obtained from (\ref{trf});

iv) linear independence of the $\L_i$ and, more strongly, the following
minimality, or ``primitivity'', condition \cite{ls}, 
\be
n_1 \L_1 +\cdots+ n_k \L_k \neq m\L, \quad {\rm with}\quad m > 1.
\label{primiti}
\ee
Solutions to this equation with integer $\L$ vector and $m> 1$
could imply spontaneous symmetry breaking of time-reversal symmetry.
For example, in the one component case, the square of the mass term
$U=(\ov\Psi^\dag \Psi)^2$ is time-reversal invariant, but it would 
induce the time-reversal breaking expectation value 
$\langle \ov\Psi^\dag \Psi \rangle \neq 0$.

The stability analysis of Refs. \cite{ls} \cite{chamon} answered the
question of whether there exist enough interactions $\L_i$ for gapping
all $N$ modes in a given Abelian theory, thus leading to a trivial
massive phase.  It was found that there always exist $(N-1)$ gapping
interactions for any matrix $\mathcal{K}$, such that one massless mode
is possible, at most. Furthermore the $N$-th gapping interaction was
show to be time-reversal invariant when the $\Z_2$ Levin-Stern index
is one, thus matching the results of the flux insertion argument. Let
us recall the main steps of this analysis.

The $(N-1)$ solutions $\L_i$ of conditions (\ref{neutral})-(\ref{primiti})
are eigenvectors of the $\S_1$ matrix
\eqref{sig1} with eigenvalue one, that can be taken of the form: 
\ba
\label{lambda} 
\L_1= \big(\L_1^{\uparrow},  \L_1^{\downarrow}\big)=
\big( \underbrace{1 ,  -1 ,  0 , \cdots,   0}_N,   
\underbrace{1,   -1,    0,  \cdots,     0}_N \big), 
\\ \vdots  \ \ \ \ \ \ \ \ \ \ \ \ \ \ \ \ \ \ \ \ \ \ \ \ \ \ 
\ \ \ \ \nonumber \\ 
\label{lambda1} 
\L_{N-1}=\big(\L_{N-1}^{\uparrow},  \L_{N-1}^{\downarrow}\big)=
\big( \underbrace{1, 0,  \cdots,    0,  -1}_N, 
\underbrace{ 1,     0,      \cdots,      0,       -1}_N   \big). 
\ea 
These vectors are globally neutral, $Q=\L_i^T\r=0$, 
but they also have neutral chiral and antichiral components, 
$\L_i^{\uparrow T} \r^\uparrow = \L_i^{\downarrow T}\r^\downarrow=0$.
As shown in Refs. \cite{ls} \cite{chamon}, 
these vectors also satisfy the other conditions, 
(\ref{Haldane})-(\ref{primiti}),
irrespectively of the form of $\mathcal{K}$ \eqref{Kmatrix}.
 
On the other hand, the $N$-th solution depends explicitly on 
$\mathcal{K}$; for simplicity, we shall consider the diagonal form 
for $\mathcal{K}$, i.e. with  $W=0$. We define the vector:
\be
\label{vbar} 
\ov{\L}=r\begin{pmatrix} K^{-1} \r^\uparrow\\
-K^{-1} \r^\downarrow \end{pmatrix}, 
\ee 
where $r$ is the smallest integer such that all components of
$\ov{\L}$ are integers.  This vector is (necessarily)
a eigenvector of $\S_1$ with  eigenvalue $-1$; it obeys all the conditions
(\ref{neutral})-(\ref{primiti}), but the integer quantity,  
\be
\label{Rindex} 
R=-\ov{\L}^T\ \S_{\downarrow} \ \r= r\, \r^{\downarrow T} K^{-1} \r^\downarrow,
\ee 
should be even for a time-reversal invariant interaction 
(cf. Eq.(\ref{trinv})).
It turns out that $R$ is equal to the Levin-Stern
quantity $2\D S=\nu^\uparrow/e^*$, characterizing the 
flux-insertion stability approach discussed in the Introduction.
The proof of this equivalence requires some assumptions on the 
form of the $K$ matrix \cite{ls}; in the cases of interest in this paper, we
use the facts that $K$ is definite positive and invariant under 
permutations of rows. Then, the fractional charge of
chiral excitations, defined by 
$Q=n^T K^{-1} \r^\uparrow$, with $n$ a $N$-dimensional integer vectors, take
minimal value $e^*$ for $n=(1,0,\dots,0)$, and this value is equal to 
the inverse of $r$ in (\ref{Rindex}). Since the filling fraction
is given by $\nu^\uparrow = \r^{\uparrow T} K^{-1} \r^{\uparrow}$, 
we finally obtain $R=2\D S$.

%-3------------------------------------------------

\section{Time-reversal interactions in the Pfaffian 
topological insulator}

\subsection{From the (331) state to the Pfaffian state} 

The Pfaffian state is the simplest non-Abelian 
quantum Hall state \cite{mr}; from the values of filling fraction and
minimal charge, we determine its stability index ($M$ is odd integer):
\be 
\nu^\uparrow=\frac{1}{M+1},\qquad 
e^*=\frac{1}{2M+2},\qquad  2\Delta S= \frac{\nu^\uparrow}{e*}=2, 
\qquad (-1)^{2\Delta S}=1. 
\label{qn-331} 
\ee 
Therefore, the Pfaffian topological insulator 
is unstable. In the following, its gapping interactions will be deduced 
from those of the ``parent''  Abelian topological state, 
the $(331)$ state defined by the $K$-matrix:
\be 
\label{k331} 
K=\begin{pmatrix} 2+M& M \\ M& 2+M\end{pmatrix}.
\ee 
(For general $M=1,3,\dots$, we should rather call it 
the $(M+2,M+2,M)$ state). 

The correspondence between $(331)$ and Pfaffian (chiral) Hall states 
is rather well established, and follows from the way
the two systems describe electrons forming bosonic
pairs \cite{cgt2}. From the $K$ matrix (\ref{k331}) we read the expression of
the $(331)$ ground state wavefunction:
\be
\Psi_{(331)}(z_i;w_j)= \prod^N_{i<j} z_{ij}^{2+M} 
\prod^N_{i<j} w_{ij}^{2+M} 
\prod^N_{i,j} (z_i-w_j)^M,
\label{331-wf}
\ee 
where $z_{ij}=z_i-z_J$, $w_{ij}=w_i-w_j$. The two sets of coordinates
$z_i$ and $w_i$, $i=1,\dots,N$, pertain to electrons that are distinct
by an additional quantum number, say isospin up and down, and the
wavefunction is antisymmetric under exchanges of coordinates of the
same kind.  Therefore, in the bosonic case $M=0$, this function does
not vanish when two electrons of opposite isospin meet at the same
point.  This feature signals the pairing of distinct electrons.

The Pfaffian state also describes electrons forming pairs with
parallel spin, ($p$-wave superconductor) and its wavefunction has the
same vanishing property.  However, in this case all the electrons are
identical and the wavefunction is completely
antisymmetric. Its expression can be obtained from the $(331)$
wavefunction by antisymmetrizing with respect to all $2N$
electron coordinates, such that the isospin quantum number is washed
out. Indeed, the following relation holds \cite{cgt2}:
\be
\Psi_{\rm Pfaff}(z_i,z_{i+n})=
{\cal A}\left[\Psi_{(331)}(z_i;w_j) \right] =
 \prod^{2N}_{i<j} z_{ij}^{M+1}\ {\rm Pf}\left( 
\frac{1}{z_i-z_j}\right),
\label{pfaff-wf}
\ee
where ${\cal A}\left[..\right]$ denotes antisymmetrization over
all the $2N$ coordinates.

The projection to identical fermions corresponds to a map 
between the conformal field theory descriptions of the two states.
In order to describe this correspondence, we should first separate 
the neutral and charged components of electron fields  (cf. (\ref{ele})).
In the $(331)$ theory, both parts are expressed by vertex operators of 
chiral bosonic fields, $\vf$ and $\phi$, that are linear combinations
of earlier field $\Phi$ (chiral part): 
\be
V=\exp\left(i\a\vf \right),\qquad F =\exp\left(i\phi \right),
\label{331-cf1}
\ee 
with $\a=\sqrt{M+1}$. The dimensions of the fields are
$h=\a^2/2=(M+1)/2$ and $h=1/2$, respectively. The field $F$ is actually
a Weyl fermion whose charge does not contribute to $Q$ but accounts
for the isospin. Thus, the charge-neutral decomposition reads:
\be
\Psi_1  = V\, F,\qquad \Psi_2=V \, F^\dagger,
\label{331-cf2}
\ee
where $\Psi_i$ have been defined earlier within the $K$-matrix lattice.

It is instructive to write the $(331)$ wavefunction as a correlator
of the bosonic fields $V$ and $F$:
\ba
\Psi_{(331)}&=&
\left\langle V(z_1)\cdots V(z_N) V(w_1)\cdots V(w_N)\right\rangle
\left\langle F(z_1)\cdots F(z_N) F^\dag(w_1)\cdots F^\dag(w_N)\right\rangle
\nl
&=&\left(\prod_{i<j} z_{ij}\, w_{ij} \prod_{i,j} (z_i-w_j)\right)^{M+1}
\ \det\left(\frac{1}{z_i-w_j} \right).
\label{331-wf2}
\ea
This expression is actually equal to (\ref{331-wf}) owing to the 
Cauchy determinant identity \cite{cft} (up to an overall constant).

The projection from the Abelian to the Pfaffian states is 
obtained by identifying the two species of Abelian
fermions $\Psi_1\sim\Psi_2$, thus eliminating the isospin
quantum number \cite{cgt2}. This amounts to projecting the Weyl fermion
to a neutral Majorana fermion, $F\to\chi$ and $F^\dag\to\chi$, namely:
\be
\Psi_1\ \to\ V\, \chi,\qquad \Psi_2\ \to\ V\, \chi.
\label{proj-1}
\ee
After this replacement, the wavefunction (\ref{331-wf2}) 
is now modified in the second term, involving the correlator
of $2N$ Majorana fields, that produces the Pfaffian expression in 
(\ref{pfaff-wf}). The clustering property of this wavefunction
 is reproduced by the fusion rule of Majorana fields $\c\cdot\c\sim I$.
The corresponding conformal
theory changes from the Abelian $U(1)\times U(1)$ to the non-Abelian
$U(1)\times \rm Ising$ theory, with central charges $c=2$ and $c=3/2$,
respectively \cite{mr}.

We now describe the corresponding map between the Abelian and
Pfaffian topological insulators.  
The Abelian insulator is defined by the ${\cal K}$ matrix
(\ref{Kmatrix}) in block-diagonal form ($W=0$), and $K$ given by (\ref{k331}).
Besides the chiral spin-up Hall states discussed so far, there 
are corresponding antichiral spin-down states, whose electrons
fields $\Psi_i$, $i=3,4$ in (\ref{ele}) are similarly 
projected into antichiral Pfaffian fields:
$\Psi_{3,4} \ \to\ \ov{V}\, \ov\chi$. 
Summarizing, we have the following map between electrons in the two
theories ($i=1,2$),
\ba
&&\Psi_{i}^{\dag} \ = \ : \exp\left(-i e_i^T \mathcal{K} \F  \right): 
\quad \ \quad
\to \quad\ :\exp\left(-i \a \vf\right): \c \ =\ V^{\dag} \, \c,
\notag\\
&&\Psi_{i} \ = \ : \exp \left(i e_i^T \mathcal{K} \F \right):\quad 
\quad \ \quad
\to \quad\ :\exp \left(i \a \vf\right):\c \ =\ V \, \c,
\notag\\
&&\Psi_{i+2}^{\dag}\ = \ : \exp \left(-i \bar{e}_i^T \mathcal{K} \F  \right): 
\quad
\ \ \to \quad\ :\exp \left(i \a \ov{\vf}\right): \ov{\c} \ =
\  \ov{V}^{\dag} \, \ov{\c},
\notag\\
&&\Psi_{i+2} \ = \ : \exp \left(i \bar{e}_i^T \mathcal{K} \F  \right):\  \quad
\quad \to \quad\ :\exp \left(-i \a \ov{\vf}\right): \ov{\c} \ =
\  \ov{V}\, \ov{\c}.
\label{proj2}
\ea
In this table, we also explain our notation for conformal fields: the bar
denotes antichirality, e.g. $\vf =\vf (z)$, $\ov{\vf} =\ov{\vf} (\bar z)$,
while the dagger refers to Fock space operators.

The time-reversal transformations of electron fields 
(\ref{tr-ele}) are left invariant by the projection, and act on the 
the electrons of the Pfaffian theory as follows:
\ba 
{\cal T}:\quad \Psi_{i}^{\dag}&=&V^\dag \, \chi \ \ \ \  \to \ \ \ \
\Psi_{i+2}^{\dag} =\ov{V}^\dag\,\ov\chi,\qquad \quad i=1,2, 
\nl 
\Psi_{i+2}^{\dag}&=&\ov{V}^\dag\,\ov\chi \ \ \ \  \to \ \ \ \ 
-\Psi_{i}^{\dag}=- V^\dag \, \chi.
\label{TR-proj}
\ea 
It would be tempting to assign the minus sign to the transformation
of the Majorana field $\c$ and leave invariant the charged Abelian field
$V$, up to antiunitarity. However, this choice is not correct and
actually could not be consistently applied to the parafermions
of Read-Rezayi states to be discussed in Section 4.2.
The correct choice is:
\ba
{\cal T}: &&V^{\dag} \ \to \ \ov{V}^{\dag}, \qquad  \quad
\ov{V}^{\dag}\ \to \ -V^{\dag},
\nl
&&\c \ \ \to \ \ov{\c}, \qquad\ \quad \   \ov{\c} \ \  \to \ \c. 
\label{tr-bosefermi} 
\ea 
These transformation rules are motivated by the following argument.
The sign is due to the $2\pi$ spin rotation of $(2+1)$-dimensional
spinors and is not built-in in the $(1+1)$-dimensional conformal
theory description.
Nevertheless, it should hinge on a conserved quantity of the conformal theory.
According to the flux-insertion argument discussed in the Introduction,
spin is represented in this theory by the difference of charges for
chiral-spin-up and antichiral-spin-down conformal fields; thus,
the fermion number sign can be written:
\be
(-1)^{N_F} =(-1)^{2S}=(-1)^{Q-\ov Q}.
\label{}
\ee 
Therefore, it is a property to be assigned to the charged fields,
and the neutral fields (Abelian and non-Abelian) transform as scalar
quantities. Note that this action of time-reversal transformations
on edge fermions is specific of topological insulators and is different
from that of topological superconductors, whose neutral
Majorana fields should carry the fermion number necessarily 
\cite{qz-rev} \cite{qz} \cite{rz}.

%-3.2--------------------------------------

\subsection{Projected interactions}

The two time-reversal-invariant interactions of the $(331)$ Abelian theory 
are obtained by the methods described in Section two; the first one
is associated to the lattice vector $\L_1$ (\ref{lambda})
 and the second one is obtained by specializing the
expression of the vector $\ov{\L}$ (\ref{vbar}) for the
$K$ matrix (\ref{k331}). They read:
\be 
\L_1=(1,-1,1,-1), \qquad \ov{\L}=(1,1,-1,-1). 
\ee 
These vectors determine  the following 
normal-ordered product of fermionic fields (cf. (\ref{H.int})):
\ba
&&U_{\L_1}=:\Psi_1^\dag\, \Psi_2\, \Psi_3^\dag\, \Psi_4: + \text{h.c.},
\nl
&&U_{\ov\L}=:\Psi_1^\dag\, \Psi_2^\dag\, \Psi_3\, \Psi_4:+ \text{h.c.},
\label{abel-inter}
\ea
(recall that the $\Psi_{1,2}$ are chiral spin-up and 
$\Psi_{3,4}$ are antichiral spin-down).
We now apply the projection to these expressions for obtaining
time-reversal invariant interactions in the
Pfaffian topological insulator. Since the maps (\ref{proj2})
apply to individual fermion fields, we should first undo the normal
ordering in (\ref{abel-inter}) by point splitting,
then apply the projection and finally re-normal order the result
in the Pfaffian theory.
Let us consider the two interactions $U_{\L_1}$ and $U_{\ov\L}$ in turn.

We use the normal-ordering formula for vertex operators \cite{cft},
\be
:\exp\left(i\a \vf(z_1)\right):\ :\exp\left(i\b \vf(z_2)\right):\, =
\, \left(z_{12}\right)^{\a \b}\ :\exp\left(i\a \vf(z_1)+i\b \vf(z_2)\right):,
\label{vertex}
\ee
to rewrite:
\be
U_{\L_1}=\lim_{z_1\to z_2} z_{12}^M\, \ov{z}_{12}^M\
\Psi_1^\dag(z_1)\, \Psi_2(z_2)\, \Psi_3^\dag(\ov{z}_1)\, \Psi_4(\ov{z}_2)
+ \text{h.c.} .
\label{int-a1}
\ee

We now perform the projection (\ref{proj2}), for each
field in this expression and then use (\ref{vertex}) to 
normal-order the vertex operators $VV^\dag$ again. We obtain:
\be
U_{\L_1}\ \to\ U_{\L_1}^{\rm Pfaff}=
\lim_{z_1\to z_2} \bigg[\frac{1}{z_{12}}
\, :V^\dag(z_1) \, V(z_2):\, \c(z_1)\, \c(z_2)\bigg]_{\rm reg.}\times
\bigg[\ z\to \bar z\ \bigg].
\label{int-a2}
\ee

Next, we consider the product expansions of chiral vertex operators and
of chiral Majorana fields for $z_1 \to z_2$, focusing on the chiral
parts. These expansions involve descendant fields in
the conformal representation (sector) of the identity field $I$ of
both the charged $c=1$ and neutral Majorana $c=1/2$ theories, owing to
the fusion rules $V^\dag\cdot V\sim I$ and $\c \cdot \c = I$: schematically,
$  U_{\L_1}^{\rm Pfaff}=\left[ I\right]_{c=1}\, \left[ I\right]_{c=1/2} $ 
\cite{cft}.
Upon using (\ref{vertex}), we find the following terms 
in the charged part (omitting constants),
\be
:V^\dag(\eps)\, V(0):= 1+\eps \de \vf + 
\eps^2 \left( (\de \vf)^2 + \de^2 \vf \right)+O(\eps^3),
\label{vertex-exp}
\ee
and in the neutral part,
\be
\c(\eps)\, \c(0) =\frac{1}{\eps} + \eps\! :\c \de\c: +O(\eps^3).
\label{fermion-exp}
\ee

The expression of $U_{\L_1}^{\rm Pfaff}$ is obtained by selecting the
finite terms for $\eps\to 0$ in the product of (\ref{vertex-exp}) and
(\ref{fermion-exp}).  This is the normal-ordering procedure for
general conformal theories to be further discussed later.  The final
expression for the interaction is obtained as follows, neglecting total
derivatives:
\be 
\label{L1e} 
U_{\L_1}^{\rm Pfaff}=\big(2T_n + \a^2\, T_c) 
\big(2\ov{T}_n + \a^2\,\ov{T}_c\big), \qquad\ \  \a^2=M+1,
\ee 
where $T_n=-\c\de \c/2$ and $T_c=-(\de \vf)^2/2$ are the stress tensors 
of the Majorana fermion and bosonic theory, respectively \cite{cft}.

In the case of the interaction $U_{\ov{\L}}$ in (\ref{abel-inter}),
we follow similar steps and arrive to the expression:
\be
U_{\ov\L}\ \to\ U_{\ov\L}^{\rm Pfaff}=
\lim_{z_1\to z_2} \bigg[z_{12}
\, :V(z_1)^\dag\, V(z_2)^\dag:\, \c(z_1)\, \c(z_2)\bigg]_{\rm reg.}\times
\bigg[\ z\to \bar{z}\ \bigg]+ {\rm h.c.}.
\label{int-b1}
\ee
The conformal sectors involved are
$  U_{\ov\L}^{\rm Pfaff}=
\left[ V^{\dag 2} \right]_{c=1}\, \left[ I\right]_{c=1/2} $,
since the original Abelian interaction had charged chiral/antichiral
parts, i.e. $Q=(2,-2)$ in the notation of Section two. 
After re-normal ordering, we finally obtain:
\be
U_{\ov\L}^{\rm Pfaff}=: \left(V^\dag\right)^2 \left(\ov{V}\right)^2 :
+ {\rm h.c.}.
\label{L2e}
\ee

Let us discuss the normal-ordering procedure employed in deriving the
non-Abelian interactions. At first sight, one could notice a certain
degree of arbitrariness in choosing the (singular) term to be
extracted from the Laurent expansions in $\eps$ (\ref{vertex-exp}),
(\ref{fermion-exp}): actually, there is none.  Each interaction is
associated to a specific representation (sector) of the conformal
theory, respectively $[I]\times [I]$ and $[V^{\dag 2}]\times [I]$,
that is selected by the quantum numbers of the electrons involved.
The operator expansions are series with coefficients the descendant
fields in these representations \cite{cft}. The normal-ordering
procedure identifies the first significant term in the series, made of
primary or descendant (quasi-primary) fields.  Had we chosen another
term of the series, its ability for gapping the system would have been
the same, since higher descendants in the representation obey the same
selection rules for coupling.

In non-Abelian theories, operator expansions may involve
more than one sector, leading to several interaction
channels, that could be selected by different normal orderings.
However, in our fermionic interactions there is always a single channel,
because the fermions are Abelian fields, as discussed in the
Introduction.

%-3.3-----------------------------------------
\subsection{Properties of non-Abelian interactions}

We discuss some features of the interactions
found in the non-Abelian theory and their ability to gap the system.
The two edge interactions $U_{\L_1}$ and $U_{\ov\L}$ just obtained are
time-reversal invariant because they involve bosonic stress tensors
$T\ov T$ and squares of the $V$ vertex operator, respectively.

The expression of $U_{\L_1}^{\rm Pfaff}$ is quartic in the Majorana
field; actually, the simpler quadratic interaction 
in the $(331)$ Abelian  theory \cite{ls},
$U= \Psi^\dag_1\, \Psi_4-\Psi_2^\dag\, \Psi_3 + {\rm h.c.}$,
can be shown to vanish after projection into the Pfaffian. 
Thus, the quartic interaction $U_{\L_1}^{\rm Pfaff}$ 
should be consider as minimal, or ``primitive'' (cf. (\ref{primiti})).

The second interaction $U_{\ov\L}^{\rm Pfaff}$ is also quartic, with
$V$ transforming under time-reversal as a single fermion in the one-component
Abelian theory. As discussed in Section two, the analogous Abelian
 interaction $\left(\Psi^\dag\Psi\right)^2$
must be discarded, because it would imply the symmetry breaking
expectation value of  the ``square-root''
$\langle\Psi^\dag\Psi\rangle \neq 0$  \cite{ls} \cite{chamon}.
In the Pfaffian theory the corresponding
time-reversal breaking quantity $V^\dag\,\ov{V}$ 
cannot acquire an expectation value because it is not
local with respect to some excitations.

For example, we consider the chiral quasiparticle $V_\s \s$ of minimal charge
$Q=1/4$ (for $M=1$), that can exist at the edge when a corresponding
quasiparticle is present in the bulk. The operator product expansion between 
the charge part $V_\s=\exp(i1/\sqrt{8}\vf)$ and $V^\dag\,\ov{V}$  is given by 
$V_\s\, V^\dag\sim z^{-1/2}$. Therefore, the following correlator 
involving two $\s$  has a square-root branch-cut and nontrivial
monodromy:
\ba
\left\langle V^\dag(0)\ov{V} (0)\,  V_\s\left(z e^{i2\pi}\right) V_\s(w)
\right\rangle 
&=& - \left\langle V^\dag (0) \ov{V}(0) \,V_\s(z) V_\s(w)
\right\rangle 
\nl
&\neq&   \left\langle V^\dag (0) \ov{V}(0)\right\rangle 
\left\langle  V_\s(z)V_\s(w)\right\rangle .
\label{monodromy}
\ea
Equation (\ref{monodromy}) implies $\langle  V^\dag\,\ov{V}\rangle=0$.
On the contrary, the Abelian operator $\Psi^\dag\Psi$ can 
acquire an expectation value
because is local with respect to all excitations.

Next, we discuss the effect of interactions  $U_{\L_1}^{\rm Pfaff}$ (\ref{L1e})
and  $U_{\ov\L}^{\rm Pfaff}$ (\ref{L2e}) within perturbation expansion.
Let us consider correlation functions of fermionic fields
$\psi=V\,\c$ and $\ov{\psi}= \ov{V}\ov{\c}$ to a given perturbative order of 
the interaction $g\,U_\L$:
\be
\label{conf-corr}
\langle \psi^\dag(z_1)\cdots \psi(z_i)\cdots\ov{\psi}^\dag(\ov{z}_j)\cdots
\ov{\psi}(\ov{z}_N)\, g \int U_\L\cdots g\int U_\L\rangle_{CFT}.
\ee
Using fusion rules and operator product expansions, we can check
whether these expressions are non-vanishing, thus
breaking conformal invariance  and eventually gapping the spectrum.

Let us first test this formula for the $N$-component Abelian theory of
Section two. In this case, the fusion of the interaction specified by $\L_k$ in
(\ref{lambda}) with the basic chiral and antichiral fermions,
respectively $\Psi_i=\Psi_i(z)$ and $\Psi_{i+n}=\ov{\Psi}_i(\ov {z})$,
$i=1,\dots,N$, takes the following form (for simplicity, consider 
${\cal  K}={\rm diag}(K,-K)$): 
\be
\label{}
\Psi^\dag_i(z)\, U_{\L_k}(0) \sim z^{\, e_i^TK\L_k^\uparrow},
\qquad\quad \ov{\Psi}^\dag_i(\ov{z})\, U_{\L_k}(0) \sim \ov{z}^{\,
  \ov{e}_i^TK\L_k^\downarrow}.  
\ee 
In presence of $n<N$ interactions
$\L_k$, we can find $(N-n)$ vectors $e_i$ (resp. $\ov{e}_i$) for which the
expression at the exponent of $z$ (resp. $\ov{z}$) vanishes, since $K$
is an invertible matrix: the electrons corresponding to these  $e_i$
and $\ov{e}_i$ do not couple to the interactions and remain massless.  
Therefore, exactly $n=N$ perturbations
$\L_k$ are needed to break scale invariance of the entire spectrum.
The argument can be repeated for quasiparticle excitations.

In Abelian theories, we further observe that the cosine interactions
(\ref{H.int}) localize the values of bosonic fields at a minimum,
leading to gapful excitations --- the so-called Haldane's
semiclassical gapping criterium \cite{ls} \cite{chamon} \cite{hal}.
Therefore, the stronger statement can be made that the $N$
interactions not only break scale invariance but give mass to all
excitations.

In the Pfaffian theory, we first distinguish between interactions
that do or do not involve charge transfer between
chiralities. Specifically, the interaction $U_{\ov\L}^{\rm Pfaff}$,
with chiral and antichiral charges $(Q,\ov Q)=(2,2)$, is non-vanishing
for the correlator,
\be
\langle \psi^\dag(z_1)\,\psi^\dag(z_2)
\ov{\psi}(\ov{z}_3)\,\ov{\psi}(\ov{z}_4)\, U_{\ov\L}^{\rm Pfaff}\rangle\neq 0.
\label{pfaff-charge}
\ee
that describes scattering with charge transfer between
the two chiralities. This process can be realized by charged 
excitations, electrons and quasiparticles, that all acquire mass by
the Haldane argument.

%-3.4.1--------------------------------------------

\subsubsection{Gapping neutral excitations}

The Pfaffian theory further possesses neutral quasiparticle excitations
that do not couple to $U_{\ov\L}^{\rm Pfaff}$. Indeed, their
scattering processes do not involve charge transfer 
and correspond to vanishing chiral correlators, as e.g.
$\langle\psi^\dag(z_1)\,\psi(z_2)\, U_{\ov\L}^{\rm Pfaff}\rangle= 0$.
Neutral excitations are affected by the other interaction 
$U_{\L}^{\rm Pfaff}$ with $(Q,\bar Q)=(0,0)$, since
$\langle \psi^\dag(z_1)\,\psi(z_2)\, U_\L^{\rm Pfaff}\rangle\neq 0$.  
In the following, we show that this coupling breaks scale invariance 
but cannot give mass to neutral excitations.

The single-edge partition function provides 
the list of quasiparticle sectors of the Pfaffian topological
insulator. As described in our earlier paper, this is obtained 
by summing the Neveu-Schwarz and Ramond sectors,
$NS, \wt{NS}, R, \wt{R}$, whose expressions are given in Appendix A.4
\cite{cr}, leading to:
\ba
\label{Pfaffian-pf}
Z_{\rm Pf\ TI} &=& Z_{NS}+ Z_{\wt{NS}} +Z_R + Z_{\wt{R}} 
\nl
&=& 2 \sum_{a=-3}^4 \left( \big| K_a I\big|^2 + \big| K_a\psi \big|^2 
+ \big| K_a \s \big|^2 \right).
\ea
In this expression, the $K_a$ are characters of $U(1)$ representations
corresponding to the Abelian parts of
excitations, carrying charge $Q=a/4+2 \Z $, while the characters $I$, $\psi$
and $\s$ describe the neutral non-Abelian parts, being the 
identity, fermion and spin of the Ising model, respectively \cite{cft}.  

In presence of the charged interaction $U_{\ov\L}^{\rm Pfaff}$ 
with large coupling, all charged excitations become highly
massive, such that $K_a \to \delta_{a,0}$ in (\ref{Pfaffian-pf})
(up to an irrelevant  factor).
Therefore, there remain the neutral excitations of the Ising model,
\be
Z_{\rm Pf\ TI} \to Z_{\rm Ising}=
 \big| I  \big|^2 + \big| \psi \big|^2 +  \big| \s \big|^2, 
\label{Ising-pf}
\ee
that are time-reversal invariant and non-chiral.

The consequences of adding the interaction 
$g U_{\L}^{\rm  Pfaff}= g T_n\ov T_n$ to the Ising model, 
with $T_n\sim \c \de\c$
its stress tensor, are well understood in the literature
\cite{zamo}.  This irrelevant interaction can be used to describe the
renormalization group flow from the tricritical to the critical Ising
model, as view from the low-energy end point. All along this flow, the
Majorana field $\chi$ stays massless, thus this interaction does not
gap the system.  A direct way to check this fact is to notice that the
derivative field present in $T$ implies a power-law correction to the
single-particle dispersion relation, $\eps (k)\sim v |k| + g
|k^3|$. Thus, values of $g\neq 0$ break conformal invariance but leave
a massless neutral state at low-energy.  This should be contrasted
with the results in the $(331)$ parent Abelian theory, where the
neutral interaction is also of cosine type and yields massive
excitations.  Note also that a $T\ov T$ interaction is possible in any
CFT consistently with time-reversal symmetry: it would imply
power-law correlations and power-law localization of neutral modes
that are irrelevant.

We now remark that the Ising model (\ref{Ising-pf}) possesses the 
relevant interaction,
\be
\label{qp-int}
U_{pq}=m\ov{\c}\c, 
\ee
that generically gives mass to the theory,
unless fine-tuning to $m=0$ is considered.
This corresponds to a quasiparticle interaction in the original
Pfaffian topological insulator, those possibility was not considered 
before. Actually, earlier discussions were limited to electron interactions,
corresponding to impurity scattering, that
are local with respect to all chiral excitations. Quasiparticle interactions,
such as $m\ov{\c}\c$, were discarded because they
can be non-local with some chiral quasiparticles.
However, in the reduced theory (\ref{Ising-pf}), 
electrons and charged chiral quasiparticles have disappeared, thus
the $m\ov{\c}\c$ interaction is local with respect to the remaining 
neutral (non-chiral) excitations and is acceptable.

In conclusion, in the Pfaffian topological insulator we introduced a
quasiparticle interaction for gapping the neutral non-Abelian modes 
that is allowed when the charged excitations are infinitely massive.
This argument requires a separation of scales between heavy charged
excitations and light neutral excitations, that is not necessary in
Abelian systems. Moreover, such quasiparticle interaction is 
generically unavoidable, but the underlying physical mechanism for its
occurrence is yet unclear. We can only speculate
that it is a sort of gravitational interaction at the edge.

We finally remark that quasiparticle interactions could be possible
in all theories, Abelian and non-Abelian, provided the charged modes
first acquire mass. The stability analysis has indeed shown that
this is the crucial fact, governed by the Levin-Stern index, and
determines all other results.

%-4-----------------------------------------

\section{Time-reversal interactions in the Read-Rezayi topological
insulators} 

\subsection{Projected interactions}

The Read-Rezayi Hall states describe the binding of identical
electrons in clusters of $k$ elements, extending the earlier $k=2$ case of
the Pfaffian \cite{rr}. The clustering implies that the ground state (bosonic)
wavefunction does not vanish when $k$ coordinates coincide.  In the
conformal field theory description, the electrons are represented by
the $\Z_k$ parafermion field $\c_1$, whose $k$-th fusion with itself 
yields the identity, $\left(\c_1\right)^k\sim I$, leading to
a non-vanishing correlator at coincident points. 
The parafermion conformal field theory is denoted by $PF_k$ and 
can be realized by the coset $PF_k=SU(2)_k/U(1)$ \cite{cv}.
As usual, excitations also have a charge part expressed 
by vertex operators, leading to the theory 
$U(1)\times PF_k$, with central charge $c=1+c_k$, $c_k=2(k-1)/(k+2)$.
The fusion of $n<k$  parafermions $\c_1$ define the parafermion field
$\c_n$, and these fields obey Abelian fusion rules among themselves:
\be
\c_i\cdot \c_j \sim \c_\ell, \qquad \ell=i+j {\rm \ mod}\ k,
\label{}
\ee
that conserve a $\Z_k$ quantum number. Moreover, they obey
$\c_n^\dag=\c_{k-n}$.

The parent Abelian theory is  a $k$-fluid
generalization of the $(331)$ state with the following $K$ matrix 
\cite{rr} \cite{cgt2}:
\be 
\label{kgen} 
K_{ij}=\left\{\begin{array}{lll}
      M+2 && i=j=1,\cdots,k, \\ 
      M  && i\neq j\, .
\end{array} \right.
\ee 
This theory describes the clustering of distinguishable electrons, 
since its wavefunction does not vanish (for $M=0$) when $k$ electrons of
different species meet at the same point. 
This wavefunction reproduce the Read-Rezayi expression upon
complete antisymmetrization with respect to all coordinates, (see
Ref.\cite{cgt2} for a complete discussion of the projection).  
The two systems share the same spectrum of charges: 
the filling fraction and minimal charge are,
\be 
\nu^\uparrow=\frac{k}{kM+2}, 
\qquad e^*=\frac{1}{kM+2}, 
\qquad  2\Delta S =k, \qquad 
(-1)^{2\D S}=(-1)^k.
\label{rr-qn} 
\ee 
Thus, the flux insertion argument tells that the topological insulators made
by pairs of Read-Rezayi states are stable (unstable) for $k$ odd (even).
The stability parameters are $(k,p)=(k,kM+2)$.

We again use the projection from the Abelian states to determine
the gapping interaction of the Read-Rezayi topological insulators.
In the present case the projection maps $k$ different chiral
species into a single one and the corresponding electron
fields $\Psi_i$, $i=1,\dots,k$ in \eqref{ele} are projected
into the Read-Rezayi electron  $\psi=V\c_1$.
More precisely, the correspondence is as follows ($i=1,\dots,k$): 
\ba 
&&\Psi_{i}^{\dag} =  : \exp \big(-i e_i^T \mathcal{K} \F  \big): \quad \ \
\quad\to \quad :\exp \big(-i \a  \vf\big): \c_1 \ 
=\ V^{\dag} \ \c_1,
\notag\\ 
&&\Psi_{i}  = : \exp \big(i e_i^T \mathcal{K} \F  \big):\quad\qquad \ \ \ 
\to \  \ \  :\exp \big(i \a  \vf\big): \c_1^{\dag} \ 
\ \ \  =V\ \c_{k-1},
\notag\\ 
&&\Psi_{i+k}^{\dag}  = 
 : \exp \big(-i \bar{e}_i^T \mathcal{K} \F \big): \ \ \ \ \
\to \ \  \ :\exp \big(i \a  \ov{\vf}\big): \ov{\c}_1 \ 
\ \  = \ov{V}^{\dag}\ \ov{\c}_1,
\notag\\ 
&&\Psi_{i+k}  = 
 : \exp \big(i \bar{e}_i^T \mathcal{K} \F \big):\ \quad \quad \ \
\to \ \ \  :\exp \big(-i \a  \ov{\vf}\big): \ov{\c}_1^{\dag}
  =\  \ov{V}\ \ov{\c}_{k-1}, 
 \label{projk} 
\ea 
where $\a^2=(2+kM)/k$ and $\vf=\vf (z),\ \ov{\vf} =\ov{\vf}(\bar z)$. 

There are $k$ Abelian gapping interactions that 
are expressed by $(k-1)$ vectors $\L_i$ and by $\ov\L$,
given in \eqref{lambda}, (\ref{lambda1})  and (\ref{vbar}), respectively.
The $\L_i$ are independent of the form of $K$; for example:
\be 
\label{1vec} 
\L_1= \big(\L_1^{\uparrow},  \L_1^{\downarrow}\big)=
\big( \underbrace{1 ,  -1 ,  0 , \cdots,   0}_k,  
 \underbrace{1,   -1,    0,  \cdots,     0}_k \big).
\ee 
The corresponding interactions have chiral components with vanishing charge.
The $k$-th vector is $\ov{\L}$ in \eqref{vbar};
for $K$ given in (\ref{kgen}), it reads:
\be 
\label{vbar331} 
\ov{\L}= \big(\ov{\L}_{\uparrow} \ ,\ 
\ov{\L}_{\downarrow}  \big)= 
\big( \underbrace{1,1,\cdots,1}_k,\underbrace{-1,-1,\cdots,-1}_k\big). 
\ee 
This vector is globally neutral because $Q=\ov{\L}^T \r=0$, but
its chiral and antichiral components are charged,
$\ov{\L}_{\uparrow}^T t=-\ov{\L}_{\downarrow}^T t=k$.
The corresponding interaction is time-reversal invariant for even $k$, 
in agreement with the index $(-1)^R= (-1)^k$ (cf. (\ref{ls-index-k})
 and (\ref{Rindex})).

The expressions of the non-Abelian interactions are obtained as follows. 
The $U_{\L_i}$ are quartic in the fermion fields as in the $k=2$ case,
Eq. (\ref{abel-inter}), and their projection follows similar steps.
After point splitting (\ref{vertex}) and projection (\ref{projk}),
one obtains the analogous of (\ref{int-a2}):
\be
U_{\L_i}\ \to\ U_{\L}^{\rm RR}=
\lim_{z_1\to z_2} \bigg[z_{12}^{-2/k}
\, :V^\dag(z_1) \, V(z_2):\, \c_1(z_1)\, \c_{k-1}(z_2)\bigg]_{\rm reg.}\times
\bigg[\ z\to \bar z\ \bigg].
\label{intRR-a2}
\ee
These interactions, for $i=1,\dots,k-1$, are all projected into the same
expression $ U_{\L}^{\rm RR}$, that involves the identity sectors 
for both the charged and neutral $\Z_k$ parafermion theories, owing to
the fusion rules $V^\dag \cdot V\sim I$ and $\c_1 \cdot \c_{k-1}\sim I$,
respectively.
The normal ordering of vertex operators is the same as in 
the Pfaffian case (\ref{vertex-exp}), with $\a^2=(2+kM)/k$.
For the parafermions we use the general operator expansion of
descendant fields in the identity sector \cite{cft},
\be 
 \label{nop-para} 
\c_1(\eps)\,\c_{k-1}(0)= \eps^{-2+2/k}+ \frac{2h_1}{c_k} \eps^{2/k} \ T_n(0)
+ :\c_1(0)\c_{k-1}(0): + \cdots ,
 \ee 
 \\ 
where $T_n$ is the stress tensor of the parafermion theory, 
$c_k$ its central charge and $h_1=(k-1)/k$ the conformal dimension of $\c_1$.
Upon combining this two operator expansions, we obtain:
\be 
\label{u1d} 
U_\L^{RR}=\bigg(\frac{2h_1}{c_k} T_n+\a ^2\, T_c \bigg)
\bigg(\frac{2h_1}{c_k} \ov{T}_n +\a ^2\, \ov{T}_c \bigg).
\ee 
This interaction takes the same $T\ov T$ form
of descendent of the identity already found in the Pfaffian case,
and fulfills the same properties. 
In particular, it cannot provide a mass for neutral excitations.

The projection of the Abelian interaction corresponding to $\ov{\L}$ is
slightly more difficult, because it involves $2k$ fermionic fields: 
\ba
\label{vbk1} 
U_{\ov{\L}}&=&:\prod_{i=1}^k\Psi^\dag_i:\ :\prod_{i=1}^k\Psi_{k+i}:
+{\rm h.c.}
\nl
&= & :\prod_{i=1}^k \exp 
\bigg(-i e_i \mathcal{K}\F(z)\bigg):\ 
: \prod_{i=1}^k \exp \bigg(i \bar{e}_i \mathcal{K}\F(z)\bigg):
 +{\rm h.c.}.
\ea 
The operators in each chiral part should be split in $k$ 
different points $\{z_1,\dots,z_k\}$, with 
$\vert z_i-z_j\vert=\eps \ \  \forall i,j$,  and later brought back to
a common point, $\eps\to 0$. 
We use the formula for the normal ordering of $k$ vertex operators \cite{cft},
\be
\label{vpk} 
\prod_{i=1}^k:\exp\left(-i e_i \mathcal{K}\F(z_i)\right): =
\prod_{i<j}^k(z_i-z_j)^M\ 
:\exp\left[-i \left(\sum_{i=1}^k e_i\right) \mathcal{K} \F(z_i) \right]: ,
\ee 
where the exponent $M$ is given by the $K$-matrix element.
Upon performing the projection (\ref{projk}) on individual fermion fields,
we re-normal order the $k$ vertex operators $V^\dag$, and obtain:
\be
U_{\ov{\L}}\to U_{\ov{\L}}^{RR}=\lim_{\eps\to 0}\left[
\prod_{i<j}^k z_{ij}^{2/k} :\left( V^\dag(z)\right)^k :\,
\prod_{i=1}^k \c_i(z_i)\right] \times \bigg[ z\to \bar z\bigg]+ 
{\rm h.c.}.
\label{u-bar}
\ee
The normal ordering of parafermion fields uses the
operator product expansions \cite{cft},
\ba
\label{opepara1} 
\c_{\ell}(z)\c_{\ell^{\prime}}(0) &\sim & z^{-2\ell \ell^{\prime}/k}\,
\c_{\ell+\ell^{\prime}}(0)+\cdots, \qquad (\ell+\ell^{\prime} < k) ,
\\
\label{opepara2}
\c_{\ell}(z)\c_{k-\ell}(0) &\sim& z^{-2\ell( k-\ell)/k}
\left( 1+ z^2 \frac{2h_{\ell}}{c_k} T_n(0) + \cdots\right),
\ea 
where $h_{\ell}=\ell(k-\ell)/k$ is the dimension of $\c_{\ell}$.
The coincidence limit of the first $(k-1)$ coordinates $z_i- z=\eps\to 0$, 
creates the parafermion field $\c_{k-1}$ with singular
behavior given by the sum of exponents in (\ref{opepara1}); for the $k$-th
limit we use (\ref{opepara2}) involving the stress tensor, and obtain,
\be
\label{intk-1}
\lim_{\eps\to 0}\prod_{i=1}^k \c_i(z_i)= \eps^{1-k} \left(1+\eps^2
\frac{2h_1}{c_k} T_n(z)\right).
\ee 
This singularity exactly cancels that coming from the product
of vertex operators in (\ref{u-bar}), leading to final result:
\be 
\label{vbk4} 
U^{RR}_{\ov{\L}}=:V^{\dag k}(z)\ov{V}^k(\bar{z})+\text{h.c.}.
\ee

%-4.2--------------------------------------------

\subsection{Properties of interactions}
 
We now discuss the features of the two edge interactions
found for the Read-Rezayi topological insulators, 
paralleling the analysis of Section 3.3.
The time-reversal transformations of Read-Rezayi fields is again inherited
from the Abelian fields (\ref{tr-ele}) through the projection (\ref{projk}),
generalizing the result (\ref{TR-proj}).
It is apparent that the minus sign cannot be assigned to the 
parafermion field $\c_1$, as it would be inconsistent with the 
fusion rule $(\c_1)^k\sim I$ for $k$ odd. As already argued, it
should be attached to the charged part, leading to the transformations:
\ba
{\cal T}: &&V^{\dag} \ \to \ \ov{V}^{\dag}, \qquad \qquad 
\ov{V}^{\dag} \quad  \to \ -V^{\dag},
\nl
&&\c_i \ \ \to \ \ov{\c}_i, \qquad\ \qquad    \ov{\c}_i \quad \to \ \ \c_i,
\label{trfiel-rr} 
\ea 
in complete analogy with (\ref{tr-bosefermi}).
It follows that the neutral interaction $U_\L^{RR}\sim \ov{T}T$ 
is time-reversal invariant for any $k$, while the charged interaction 
$U_{\ov\L}^{RR}$ is only invariant for $k$
even, as in the parent Abelian theory. In the even $k$ case, we should
again consider the possibility that $U_{\ov\L}^{RR}$ breaks spontaneously
the symmetry by forcing an expectation value for $V^\dag \ov{V}$.
This cannot happen because $V^\dag \ov{V}$   is non-local with
respect to some excitations of the system. For example, 
we consider the chiral quasiparticle of smallest charge 
$e^*$ in (\ref{rr-qn}) inserted in the analogous of the correlator 
(\ref{monodromy}) of the Pfaffian case; one needs the operator expansion 
of $V^\dag$ with the quasiparticle vertex operator
$V_{e^*}=\exp(i\b \vf)$, with $\b=1/(2\sqrt{M+1})$, which is
$V^\dag(\eps)\, V_{e^*}(0)\sim \eps^{-1/\sqrt{2k}}$, leading to a non-trivial
monodromy incompatible with spontaneous symmetry breaking.
 
The properties of interactions should be compared with the analysis
based on flux insertions and Kramers theorem \cite{cr}.  For $k$ odd,
this tells that some edge excitations are protected against any
time-reversal invariant interaction and remain massless. Indeed, these
are the charged excitations, because the interaction $U_{\ov\L}^{RR}$
(\ref{u1d}) is forbidden by time-reversal invariance.

For $k$ even, the Read-Rezayi topological insulators are unstable
according to the flux argument. The charged interaction is allowed and
gaps all charged excitations, being a cosine of the Abelian field.
Regarding the other interaction, $U_{\L}^{RR}\sim \ov{T}T$, it is not
sufficient to give mass to the remaining neutral excitations, owing to the
arguments discussed in Section 3.3 for $k=2$.

In order to gap the neutral modes, we consider a quasiparticle
interaction that is allowed in the reduced neutral
theory where all charged states have acquired very large masses.  In
this limit, the single-edge partition function of Read-Rezayi
topological insulators \cite{cr} becomes the following
expression: 
\be
\label{RR-pf}
Z^{RR}\ \to\ \sum_{\ell=0,\rm \ even}^k \left\vert \c_0^\ell\right\vert^2 +
 \sum_{\ell=0,\ \ell-\frac{k}{2}\ \rm even}^k 
\left\vert \c_{\frac{k}{2}}^\ell\right\vert^2,
\ee
that contains a subset of the excitations of the $Z_k$ parafermion
statistical model \cite{cv}. The parafermionic characters 
$\c_m^\ell$ describe neutral excitations with quantum numbers 
$(\ell,m)\equiv (\ell,m+2k)\equiv (k-\ell,m+k)$ and $m=\ell$ mod $2$.

In this theory we introduce the quasiparticle interaction,
\be
\label{rr-int}
U^{RR}_{qp} = \ov{\c}^2_0 \c_0^2 + {\rm h.c.},
\ee
with dimension $2h_0^2=4/(k+2)<2$, that is relevant and
couples to all sectors, since the fusion rules $\c^2_0\c^\ell_0$ and 
$\c^2_0\c^\ell_{\frac{k}{2}}$ are different from zero for any allowed value
of $\ell$ \cite{cft}. 
This interaction is generically nonvanishing and gives mass to all
the neutral interactions; indeed, such a relevant term
coupling to all sectors drives the system into a completely massive phase. 

In conclusion, the analysis of interactions in the Read-Rezayi
topological insulators confirms the result of the flux argument: some
charged excitations remain massless for odd $k$ values, while all
excitation become massive for even $k$.  

%-5-------------------------------------------

\section{Time-reversal interactions in the non-Abelian Spin-Singlet
state}

\subsection{The NASS state and its parent Abelian state}

In this section, we extend our analysis to another prominent
non-Abelian state involving spinful chiral electrons, 
that could be used again to model topological insulators.
This provides another test of our approach of projected Abelian
states. Some background on the physical motivations and properties
of the NASS Hall state can be found in Refs. \cite{nass}; we
 follow the analysis of the quantum numbers, spectrum and 
partition function of Ref. \cite{cv}.

In the NASS state, the clustering property of the Read-Rezayi
states is extended to spinful electrons, requiring that the (bosonic)
ground state wavefunction does not vanish when $k$ electrons
with spin up and $k$ with spin down meet at the same point.
The conformal field theory description involves generalized parafermions
of two kinds, $\c_{\uparrow 1}$ and $\c_{\downarrow 1}$, 
that obey the fusion rules $(\c_{\uparrow 1})^k=(\c_{\downarrow 1})^k=I$
and thus reproduce the vanishing property of the wavefunction.
These parafermions can be obtained from the coset theory 
$\wh{SU(3)}_k/\wh{U(1)}^2$, to which we should add the $\wh{U(1)}^2$
symmetry for charge and spin conservation.
Although this state involves up and down spin configurations,
it is chiral; thus, a time-reversal invariant
topological insulator is obtained by 
adding  a specular antichiral state, as in 
earlier spin-polarized cases. The combined system
possesses independent spin and chiral excitations of both kinds,
namely chirality and spin are not tied together. Therefore,
excitations are always doubled and a single Kramers pair cannot be
formed, leading to instability according to the flux insertion argument.
The charge parameters are:
\be
\nu^\uparrow=\frac{2k}{2kM+3}, \qquad
e^*=\frac{1}{2kM+3},\qquad (-1)^{2S}=(-1)^{2k}=1, \quad M \ {\rm odd},
\label{nass-qn}
\ee 
showing that the stability index is always one, as expected. 
Thus, we should exhibit the gapping
interactions consistent with time-reversal symmetry.

The NASS state also possesses a parent Abelian state that describes
distinguishable electrons with the same physical features, and that 
reduces to the non-Abelian state upon projection to identical
electrons \cite{nass}. 
Let us recall this Abelian theory in the simplest case $k=2$ and
charge parameter $M$ eventually fixed to $M=1$, 
leading to $\nu^\uparrow=4/7$ and $e^*=1/7$ in (\ref{nass-qn}).
For $k=2$ there are two chiral electrons for each spin orientation,
$\uparrow,\downarrow$, 
that are distinct by the index $a=1,2$, 
leading to the four-dimensional chiral $K$ matrix:
\be
\label{KSS}
K=
\begin{pmatrix} 
2 &1 &0 &0 \\
1 &2 &0 &0 \\
0 &0 &2 &1 \\
0 &0 &1 &2
\end{pmatrix}
+ M 
\begin{pmatrix} 
 1 &1 &1 &1 \\
 1 &1 &1 &1 \\
 1 &1 &1 &1 \\
 1 &1 &1 &1 \\
\end{pmatrix} .
\ee
The corresponding wavefunctions is: 
\be
\Psi^{NASS}_{\rm Ab}= \prod_{i<j} \left(
z_{ij}^\uparrow\, z_{ij}^\downarrow\, w_{ij}^\uparrow\, w_{ij}^\downarrow
\right)^2 \prod_{i,j}(z_i^\uparrow -z_j^\downarrow)(w_i^\uparrow -w_j^\downarrow),
\label{nass-apf}
\ee
where $z_i^\uparrow,z_i^\downarrow $ and $w_j^\uparrow,w_j^\downarrow$ 
are the coordinates of the two kinds of fermions,
with both spin orientations. The overall factor
$\prod_{i<j} x_{ij}^M$ has been omitted in $\Psi^{NASS}_{\rm Ab}$, 
where $x_i$ stands for any coordinate type.

The topological insulator is obtained by adding four antichiral
components, leading to a eight-dimensional ${\cal K}$ matrix in block
diagonal form.  The elementary fermion fields are vertex operators
written in terms of a eight-dimensional scalar field $\Phi$ and 
lattice basis $e_i$, $i=1,\dots,8$ defined in (\ref{ele}).
From the wavefunction (\ref{nass-apf}), the vectors of chiral electrons with
up and down spins are identified as follows:
\ba
\label{cele-ss}
&&e_1^{\uparrow}=e_1=(1,0,0,0,0,0,0,0),\quad\quad\ \ 
e_1^{\downarrow}=e_2=(0,1,0,0,0,0,0,0),\notag\\
&&e_2^{\uparrow}=e_3=(0,0,1,0,0,0,0,0),\quad\quad \ \ 
e_2^{\downarrow}=e_4=(0,0,0,1,0,0,0,0),
\ea
while for antichiral electrons, they read,
\ba
\label{acele-ss}
&&\bar{e}_1^{\downarrow}=e_5=(0,0,0,0,1,0,0,0),\quad\quad\ \ 
\bar{e}_1^{\uparrow}=e_6=(0,0,0,0,0,1,0,0),\notag\\
&&\bar{e}_2^{\downarrow}=e_7=(0,0,0,0,0,0,1,0),\quad\quad\ \ 
\bar{e}_2^{\uparrow}=e_8=(0,0,0,0,0,0,0,1).
\ea

Owing to the spin assignments in (\ref{cele-ss}) and (\ref{acele-ss}),  
we should permute the elements of the basis to match the standard notation of
Section two, in which the first (resp. second) half of the components 
describe spin up (down) electrons, and time-reversal 
transformations acts by the matrix $\S_1$. The elements of the
new basis are ordered as follows:
\be
e_1^{\uparrow}\to e_1,\ \bar{e}_1^{\uparrow}\to e_2,\ 
e_2^{\uparrow}\to e_3, \ \bar{e}_2^{\uparrow}\to e_4, \
\bar{e}_1^{\downarrow}\to e_5, \ e_1^{\downarrow}\to e_6,\
\bar{e}_2^{\downarrow}\to e_7, \ e_2^{\downarrow}\to e_8 .
\label{mapp}
\ee
In the new basis the $\mathcal{K}$ matrix is no longer block-diagonal
since the chiralities are mixed. It reads (hereafter we take $M=1$):
\be
\mathcal{K}=
\begin{pmatrix}
3 & 0 &1&0&0&2&0&1 \\
0&-3&0&-1&-2&0&-1&0\\
1&0&3&0&0&1&0&2\\
0&-1&0&-3&-1&0&-2&0\\
0&-2&0&-1&-3&0&-1&0 \\
2&0&1&0&0&3&0&1\\
0&-1&0&-2&-1&0&-3&0\\
1&0&2&0&0&1&0&3
\end{pmatrix}.
\ee
Note that time-reversal symmetry invariance,
$\mathcal{K}= - \S_1\ \mathcal{K} \ \S_1$, is verified.
The electron operators and their transformation rule under time reversal
are, as usual,
\ba
\label{TR-nass}
{\cal T}:&& \Psi_{a \uparrow}^{\dag}=
\exp\left(-ie^T_{a \uparrow}{\cal K}\Psi\right) \qquad
\to  \qquad
\ov{\Psi}_{a \downarrow}^{\dag}=
\exp\left(-i\ov{e}^T_{a \downarrow}{\cal K}\Psi\right),
\ \ \qquad a=1,2,
\notag\\
&&\Psi_{a \downarrow}^{\dag} \ \to\ -\ov{\Psi}_{a \uparrow}^{\dag},\qquad \ \ \ 
\ov{\Psi}_{a \uparrow}^{\dag} \ \to\ \Psi_{a \downarrow}^{\dag},\qquad\ \ 
\ov{\Psi}_{a \downarrow}^{\dag} \ \to\ -\Psi_{a \uparrow}^{\dag}.
\ea

The projection from the Abelian to the NASS theory is realized
by identifying electrons of the two species, as follows:
\ba
\label{proj-nass1}
&&\Psi_{a \uparrow}^{\dag} \ \ \  \to \ \ \ \ 
\c_{\uparrow} : \exp \big( -i \a \vf_{c} \big)\exp 
\big( -i \b \vf_{s} \big):\ \ =\ 
\c_{\uparrow} \ V^{\dag}_{\a}V^{\dag}_{\b}, \quad\ \ \ a=1,2,
\notag\\
&&\Psi_{a \downarrow}^{\dag}\ \ \  \to \ \ \ \ 
\c_{\downarrow}: \exp \big( -i \a \vf_{c} \big)\exp \big( i \b \vf_{s} \big):
\ \ \ \ \ \ =\ \c_{\downarrow}\ V^{\dag}_{\a}V_{\b},
\notag\\
&& \ov{\Psi}_{a \downarrow}^{\dag}\ \ \  \to \ \ \ \ 
\ov{\c}_{\downarrow} : \exp \big( i \a \ov{\vf}_{c} \big)
\exp \big(- i \b \ov{\vf}_{s} \big):
\quad \ \ \  =\ \ov{\c}_{\downarrow}\ \ov{V}^{\dag}_{\a} \ov{V}_{\b},
\notag\\
&&\ov{\Psi}_{a \uparrow}^{\dag}\ \ \  \to \ \ \ \ 
\ov{\c}_\uparrow : \exp \big( i \a \ov{\vf}_{c} \big)
\exp \big( i \b \ov{\vf}_{s} \big):
\qquad\ \ \ \  =\ \ov{\c}_\uparrow \ \ov{V}^{\dag}_{\a} \ov{V}^\dag_{\b},
\ea
where $\vf_c\ (\ov{\vf_c})$ and $\vf_s\ (\ov{\vf_s})$  are the chiral 
(antichiral) scalar fields for the charged and spin parts, respectively, and
$\c_{\uparrow}, \c_{\downarrow}$ ($\ov{\c}_{\uparrow}, \ov{\c}_{\downarrow}$) 
are the chiral (antichiral) spinful $k=2$ parafermions with conformal
dimension $h=1/2$, obeying $\c^{\dag}_\s= \c_\s$, $\s=\uparrow,\downarrow$. 
The  parameters in the vertex operators for charge $V_\a$ and
and spin $V_\b$ are $\a^2=7/4$ and $\b^2=1/4$, respectively.

The time-reversal transformations pass through the
projection to the NASS conformal fields; as in earlier cases,
we should assign the fermion sign to a conserved quantity of the
conformal theory that is related to spin. In the NASS case, the field
$V_\b$ carries the $U(1)_S$ spin symmetry so the sign can be attached 
to it as $(-1)^{2S}$. As discussed in \cite{cr},
spin is not conserved in realistic models of topological insulators,
but is broken to spin parity, $U(1)_S \to \Z_2$, 
in presence of time-reversal symmetry; thus, the sign is well-defined
in general.
In conclusion, the time-reversal transformations are:
\be
{\cal T}:V^{\dag}_{\b} \ \to \ \ov{V}_{\b}, \qquad \qquad
\ov{V}_{\b}\ \to \ -V^{\dag}_{\b},
\label{TR-spin} 
\ee
while $V_\a$ and the $\c_\s$  are only conjugated.

%-5.2--------------------------------------

\subsection{Projected interactions}

In the new basis (\ref{mapp}), the vectors specifying the
time-reversal invariant interactions of the parent Abelian theory
(\ref{lambda})-(\ref{lambda1}) and (\ref{vbar}) have the form,
according to the discussion in Section two,
\ba
\label{ss-in}
&&\L_1=(1,-1,0,0,1,-1,0,0), \qquad\ \L_2=(1,0,-1,0,1,0,-1,0), 
\notag\\
&&\L_3=(1,0,0,-1,1,0,0,-1),\ \qquad\ \ov{\L}=(1,-1,1,-1,-1,1,-1,1).
\ea
We recall that the first three obey  $\S_1\L_i=\L_i$, $i=1,2,3$,
while the forth fulfills $\S_1\ov{\L}=-\ov{\L}$, and is
time-reversal invariant owing to $\ov{\L}^T \S_{\downarrow}\r=0$.
Using the general formula for vertex operators (\ref{ele}) and keeping
in mind the identifications of the basis (\ref{mapp}), we obtain
the expressions:
\ba
\label{nass-aint}
&&U_{\L_1}= :\Psi^\dag_{1\uparrow}\, \ov{\Psi}_{1\uparrow}\, 
\ov{\Psi}^\dag_{1\downarrow}\, \Psi_{1\downarrow} :+{\rm h.c.},
\nl
&&U_{\L_2}= :\Psi^\dag_{1\uparrow}\, \Psi_{2\uparrow}\, 
\ov{\Psi}^\dag_{1\downarrow}\, \ov{\Psi}_{2\downarrow} :+{\rm h.c.},
\nl
&&U_{\L_3}= :\Psi^\dag_{1\uparrow}\, \ov{\Psi}_{2\uparrow}\, 
\ov{\Psi}^\dag_{1\downarrow}\, \Psi_{2\downarrow} :+{\rm h.c.},
\nl
&&U_{\ov\L}\, = :\Psi^\dag_{1\uparrow}\, \ov{\Psi}_{1\uparrow}\, 
\Psi^\dag_{2\uparrow}\, \ov{\Psi}_{2\uparrow}\, 
\ov{\Psi}^\dag_{1\downarrow}\, \Psi^\dag_{1\downarrow} 
\ov{\Psi}^\dag_{2\downarrow}\, \Psi^\dag_{2\downarrow} :+{\rm h.c.}.
\ea

We now compute the projected forms of each interaction  by the
method of point splitting and re-normal ordering already applied before.
The operator product expansions of Abelian electrons needed in the
calculations can be read from the wavefunction (\ref{nass-apf}):
\ba
&&\Psi^\dag_{a \uparrow}(z) \Psi^\dag_{a \downarrow}(0) \ = z^2 
:\Psi^\dag_{a \uparrow}\, \Psi^\dag_{a \downarrow}:,
\qquad\qquad a=1,2,
\nl
&& \Psi^\dag_{1 \s}(z) \Psi^\dag_{2 \s'}(0) = z
: \Psi^\dag_{1 \s}\, \Psi^\dag_{2 \s'}:,
\qquad\qquad \s,\s'=\uparrow,\downarrow,
\label{ope-nass-v}
\ea
while those of the vertex operators
$V_{\a}$ and $V_{\b}$ are expressed by the charges
$\a$ and $\b$ defined before.
We also need the following operator products of parafermions $\c_\s$, 
$\s=\uparrow,\downarrow$ \cite{nass},
\ba
\label{ope-nass-para}
&&\c_\s(z)\c_\s(0) \ \quad=\  \ z^{-1} I +\l\, z\, T_n(0)+\cdots,
\qquad \s=\uparrow,\downarrow,
 \\
&& \c_\uparrow(z)\c_\downarrow(0)\quad \ = \ \ z^{-1/2}\c_{\uparrow\downarrow}(0)-
z\,\c_\uparrow\de\c_\downarrow(0)+\cdots,
\\
&& \c_{\uparrow\downarrow}(z)\c_{\uparrow\downarrow}(0)\ \  = \ \  z^{-1}\, 
I+\cdots, 
\\
&& \c_{\uparrow\downarrow}(z)\c_{\uparrow}(0)\quad=\ \  z^{-1/2}\,
\c_{\downarrow}(0)+
\cdots,
\\
&& \c_{\uparrow\downarrow}(z)\c_{\downarrow}(0)\quad =\ \  z^{-1/2}\,
\c_{\uparrow}(0)+
\cdots,
\ea
where $\l$ is a constant, $\c_{\uparrow\downarrow}$ is 
another parafermionic field with conformal dimension $h=1/2$, and 
$T_n$ is the stress tensor of the parafermion theory.

Let us first consider the interaction $U_{\L_1}$ in (\ref{nass-aint}).
Upon splitting the two chiral fermions in points $z_1$ and $z_2$,
performing the projection and re-normal ordering the $V$ fields,
we obtain:
\ba
U_{\L_1}&\to& U_{\L_1}^{NASS} 
\nl
&=& \left[
z_{12}^{1/2}: V_\a^\dag V_a::V^\dag_\b V^\dag_\b :
\left(z_{12}^{-1/2} \c_{\uparrow\downarrow} +O(z_{12})\right)\right]
\times\Big[z\to\bar{z}, \uparrow\to\downarrow\Big]+{\rm h.c.}
\nl
&=& :V_\b^{\dag 2}\, \ov{V}_\b:\,
\c_{\uparrow\downarrow}\ov{\c}_{\uparrow\downarrow}+{\rm h.c.},
\label{U1-nass}
\ea
showing that this interaction involves the spin sector of the theory.

Next we observe that the interactions $U_{\L_3}$ and $U_{\L_1}$
differ in the type $a=1,2$ of some fields, that is irrelevant 
after projection to the NASS theory.
Nevertheless, a different singularity of original Abelian fields,
namely $\Psi^\dag_{1 \uparrow}\Psi_{1\downarrow}\sim z^{-2}$ versus
$\Psi^\dag_{1 \uparrow}\Psi_{2\downarrow}\sim z^{-1}$
(cf. (\ref{nass-apf})), implies a slightly different
result in the normal-ordering procedure :
\be
\label{U3-nass}
U_{\L_3}^{NASS} =
\left(\a \de\vf_c -\b \vf_s\right)
\left(\a \de\ov{\vf}_c -\b \ov{\vf}_s\right)
:V_\b^{\dag 2}\, \ov{V}_\b:\,
\c_{\uparrow\downarrow}\ov{\c}_{\uparrow\downarrow}+{\rm h.c.}.
\ee
Therefore, $U_{\L_3}^{NASS}$ differs from $U_{\L_1}^{NASS}$ for the 
presence of descendant fields in the same conformal sector.
Although their explicit expressions are different, we should not consider 
the two interactions as independent, because they are equivalent in the 
 ability of gapping  excitations, as discussed in Section 3.2.
 
The analysis of $U_{\L_2}$ follows similar steps and we find:
\ba
U_{\L_2}&\to& U_{\L_2}^{NASS} 
\nl
&=& \left[
z_{12}^{-1}: V_\a^\dag V_a::V^\dag_\b V_\b :
\left(z_{12}^{-1} +\l z_{12} T_{pf}(z_2)\right)\right]
\times\Big[z\to\bar{z}, \uparrow\to\downarrow\Big]+{\rm h.c.}
\nl
&=& \left(\a^2 T_c+\b^2 T_s+\l T_{pf} -\a \b \de\vf_c \de\vf_s \right)
\nl
&&\ \ \times
\left(\a^2\ov{T}_c+\b^2\ov{T}_s+\l \ov{T}_{pf} -
\a \b \bar{\de}\bar{\vf}_c \bar{\de}\bar{\vf}_s \right),
\label{U2-nass}
\ea
where $T_s=-1/2 (\de \vf_s)^2$ and $T_c=-1/2 (\de \vf_c)^2$.
It is apparent that this interaction involves descendant fields
in the identity sectors of charge, spin and parafermionic
parts of the conformal theory.

We finally consider the projection of the interaction $U_{\ov\L}$ 
in (\ref{nass-aint}).
In this case, we should separate four chiral Abelian fields at
points $\{z_1,z_2,z_3,z_4\}$, with $\vert z_i-z_j\vert=\eps$,
and then bring them back together after projection. 
Focusing on the chiral part, and using (\ref{ope-nass-para}) for fusing
 parafermions, we obtain:
\ba
U_{\ov\L}\bigg\vert_z&= &\lim_{\eps\to 0} 
\left(z_{12}z^2_{13}z_{14}z_{23}z^2_{24}z_{34}\right)^{-1} \,
\Psi^\dag_{1\uparrow}(z_1)\Psi^\dag_{2\uparrow}(z_2) 
\Psi^\dag_{1\downarrow}(z_3) \Psi^\dag_{2\downarrow}(z_4) 
\nl
&\to &\lim_{\eps\to 0} z_{12} z^{-1/2}_{13}z_{14}^{1/2}z_{23}^{1/2}z_{24}^{-1/2}z_{34}
: V^{\dag 4}_\a(z): 
\c_\uparrow(z_1) \c_\uparrow(z_2) \c_\downarrow(z_3) \c _\downarrow(z_4)
\nl
&=& \lim_{\eps\to 0}\left(\frac{z_{14} z_{23}}{z_{13}z_{24}}\right)^\frac{1}{2}
: V^{\dag 4}_\a(z): .
\ea
Therefore, the forth interaction is:
\be
\label{lbar4}
U_{\ov{\L}}^{NASS}=:V^{\dag 4}_{\a}\ov{V}^4_{\a}: +\text{h.c.}.
\ee

%-5.3----------------------------------
\subsection{Properties of NASS interactions}

Let us summarize the three independent time-reversal invariant
electron interactions that have been obtained in the NASS state with $k=2$:
\ba
 U_{\L_2}^{NASS}& =&
\left(\a^2 T_c+\b^2 T_s+\l T_{pf} -\a \b \de\vf_c \de\vf_s \right)
\nl
&&\ \ \times \left(\a^2\ov{T}_c+\b^2\ov{T}_s+\l \ov{T}_{pf} -
\a \b \bar{\de}\bar{\vf}_c \bar{\de}\bar{\vf}_s \right),
\\
 U_{\L_1}^{NASS}& =&
:V_\b^{\dag 2}\, \ov{V}_\b^2:\,
\c_{\uparrow\downarrow}\ov{\c}_{\uparrow\downarrow}+{\rm h.c.},
\\
U_{\ov{\L}}^{NASS}&=&:V^{\dag 4}_{\a}\ov{V}^4_{\a}: +\text{h.c.}.
\label{}
\ea
These interactions involve neutral, spinful and charged operators
within each chirality, respectively.  As in previous theories, we can
argue that spontaneous breaking of time-reversal symmetry cannot be
induced by $U_{\L_1}^{NASS}$, because the non-invariant operator
$V^{\dag}_\b\ov{V}_\b$ cannot acquire an expectation value, being
nonlocal with respect to some chiral excitations.
Therefore, the two interactions $U_{\L_1}^{NASS}$ and $U_{\ov{\L}}^{NASS}$
are of Abelian type and provide a mass to all charged and spinful excitations.

Regarding the neutral interaction $U_{\L_2}^{NASS}$, this involves
derivative operators and does not gap the neutral excitations,
owing to the argument of Section 3.3.  Again, we consider
the reduced theory where all charged and spinful excitations have decoupled:
in this limit, the NASS partition function \cite{cr} reduces to
the following expression:
\be
\label{NASS-pf}
Z^{NASS} \ \to\ \left\vert I \right\vert^2 +
\left\vert \rho \right\vert^2 + \left\vert \psi_{12} \right\vert^2 + 
\left\vert \s_3 \right\vert^2.
\ee
This tells us the remaining neutral quasiparticles: the  $\rho$, $\psi_{12}$
and $\s_3$ excitations of the $SU(3)$ parafermion theory \cite{nass}.
In this theory, the quasiparticle interaction,
\be
\label{NASS-qp}
U^{NASS}_{qp} = \ov{\r}\r ,
\ee
is relevant, $2h_\r=6/5<2$, and couples to 
all neutral excitations in (\ref{NASS-pf}) \cite{nass}.
Therefore, this interaction is generically present and 
drives the system into a completely massive phase. 

In conclusion, we found the needed interactions that confirm
the instability of this system as predicted by the flux-insertion 
argument.

%-6-------------------------------------

\section{Conclusions}
In this paper, we have found the edge interactions that can gap
time-reversal invariant topological insulators made by
chiral-antichiral pairs of non-Abelian Hall states.  In the cases
where the flux-insertion argument predicted the instability of the
systems, we find a sufficient set of interactions that actually let
them decay. In case of stability, as e.g. the $\Z_k$ parafermionic
Read-Rezayi states with $k$ odd, we found instead that the available
interactions are not sufficient to gap the system completely.  These
results  complement the stability analysis of our previous
paper \cite{cr}.

That unstable topological insulators possess enough interactions for
decaying could be considered as a natural result.  However, in
checking this property explicitly we found some interesting features
of these models.  Firstly, we noticed that the spin sign in
time-reversal transformations is carried by Abelian conformal fields
that can represent faithfully the spin parity $(-1)^{2S}$. Secondly,
we observed that the expressions of interactions require normal
ordering and that this procedure is free from ambiguities due to the
 electron field obeying Abelian fusion rules
(a so-called simple current) \cite{cv}.  
Thirdly, we found that neutral excitations are gapped by a
  quasiparticle interaction that is allowed and generically present in
  the low-energy limit of the theory. 
The physical origin of this interaction remains to be understood, since
it does not correspond to electron scattering at impurities. 

Our study of interaction was based on previous analyses of Abelian
theories combined with particular projections that relate Abelian and
non-Abelian Hall states \cite{cgt2} \cite{nass}. It is likely that
this kind of approach could be extended to other non-Abelian states
that are described by the more general Wen's parton construction also
involving Abelian states \cite{ringel} \cite{wen-na}.

{\bf Acknowledgments}

We would like to thank D. Bernard, T. H. Hansson, 
K. Schoutens, S. H. Simon, P. Wiegmann for interesting discussions.
We thank in particular A. Stern for useful criticism on the 
first version of the paper.
This work was partially supported by the European IRSES network,
``Quantum Integrability, Conformal Field Theory and Topological
Quantum Computation'' (QICFT).

%-B-------------------------------------------- 


\begin{thebibliography}{99} 
\bibitem{qz-rev} 
  X.~L.~Qi, S. C. Zhang, 
  ``Topological insulators and superconductors'' 
  Rev. Mod. Phys. {\bf83} (2011) 1057; 
  E. Fradkin, 
  Field Theories of Condensed Matter Systems, 2nd edition, 
  Cambridge Univ. Press (2013), Cambridge UK. 
\bibitem{molen} M.~K\"onig, S.~ Wiedmann, C.~ Bruene, A.~ Roth, H.~ 
  Buhmann, L.~ W.~ Molenkamp, X.~L.~ Qi, S.~C. Zhang, 
  ``Quantum Spin Hall Insulator State in HgTe Quantum Wells'', 
  Science {\bf 318}  (2007) 766; 
  M.~Koenig, H.~ Buhmann, L.~ W.~ Molenkamp, T.~ Hughes, C.~-X.~ Liu, 
  X.~-L.~ Qi, S.~C. Zhang, 
  ``The Quantum Spin Hall Effect: Theory and Experiment'', 
  J. Phys. Soc. Jpn. {\bf 77} (2008)  031007. 
\bibitem{3d-ti} 
  D. Hsieh, D. Qian, L. Wray, Y. Xia, Y. S.  Hor,  R. J. Cava, M. Z. Hasan, 
  ``A topological Dirac insulator in a quantum spin Hall phase'', 
  Nature {\bf 452} (2008) 970; 
  Y. Xia, D. Qian, D. Hsieh, L. Wray, A. Pal, H. Lin, A. Bansil, 
  D. Grauer, Y. S. Hor, R. J. Cava, M. Z. Hasan, 
  ``Observation of a large-gap topological-insulator class with 
  a single Dirac cone on the surface'', 
  Nature Physics {\bf 5} (2009) 398; 
  Y. L. Chen, J. G. Analytis, J.-H. Chu, Z. K. Liu, S.-K. Mo, 
  X. L. Qi, H. J. Zhang, D. H. Lu, X. Dai, 
  Z. Fang, S.~Zhang, I. R. Fisher, Z. Hussain, Z.-X. Shen, 
  ``Experimental Realization of a Three-Dimensional Topological Insulator 
  Bi$_2$Te$_3$'', 
  Science {\bf325} (2009) 178.
  \bibitem{3d-interact} 
  M. A. Metlitski, C. L. Kane, M. P. A. Fisher, 
   ``A symmetry-respecting topologically-ordered surface phase of 3d electron 
  topological insulators", 
  preprint arXiv:1306.3286; 
  X. Chen, L. Fidkowski, A. Vishwanath, 
  ``Symmetry Enforced Non-Abelian Topological Order at the 
  Surface of a Topological Insulator", Phys. Rev. B {\bf 89} (2014) 165132; 
  P. Bonderson, C. Nayak, X.-L. Qi, 
  ``A Time-Reversal Invariant Topological Phase at the Surface 
  of a 3D Topological Insulator'', 
  J. Stat. Mech. (2013) P09016.
  \bibitem{ls} 
  M.~Levin, A.~Stern, 
  ``Fractional Topological Insulator'', 
  Phys. Rev. Lett. {\bf103} (2009) 196803, 
  ``Classification and analysis of two dimensional abelian 
  fractional topological insulators,'' 
  Phys. Rev. B {\bf 86} (2012) 115131. 
   \bibitem{wen} 
  X.~Chen, Z.~X.~Liu, X.~G.~Wen, 
  ``Two-dimensional symmetry-protected topological orders and their 
  protected gapless edge excitation'', 
  Phys. Rev. B {\bf 23} (2011)  235141. 
\bibitem{chamon} 
  T.~Neupert, L.~Santos, S.~Ryu, C.~Chamon, C.~Mudry, 
  ``Fractional topological liquids with time-reversal symmetry and their 
  lattice realization'', 
  Phys. Rev. B  {\bf84} (2011) 165107; 
  ``Time-reversal symmetric hierarchy of fractional incompressible liquids'', 
  Phys. Rev. B {\bf84} (2011) 165138. 
\bibitem{vish} 
Y.~M.~Lu, A.~ Vishwanath, 
  ``Theory and classification of interacting integer topological 
  phases in two dimensions: A Chern-Simons approach'', 
  Phys. Rev. B {\bf 86} (2012) 125119.
   \bibitem{cr} A. Cappelli, E. Randellini, 
``Partition Functions and Stability Criteria of Topological Insulators'', 
JHEP 12 (2013) 101. 
 \bibitem{ringel}  Z.~ Ringel, A.~Stern 
  ``The $\mathbb{Z}_2$-anomaly and boundaries of topological insulators'', 
  Phys. Rev. B {\bf 88} (2013) 115307;  M Koch-Janusz, Z. Ringel, 
``Interacting and fractional topological insulators via the 
${Z}_{2}$ chiral anomaly", 
Phys. Rev. B {\bf 89} (2014) 075137. 
 \bibitem{ryu} 
O.~M.~Sule, X.~Chen, S.Ryu, 
``Symmetry-protected    topological phases and orbifolds'', 
Phys. Rev. B {\bf 88} (2013)   075125; 
C.T. Hsieh, O. M. Sule, G. Y. Cho, S. Ryu, R. G. Leigh,
``Symmetry-protected Topological Phases, Generalized Laughlin
   Argument and Orientifolds'', 
preprint arXiv:1403.6902; 
T. C. Hsieh,  T. Morimoto, S. Ryu, 
``CPT theorem and classification of topological insulators 
and superconductors'', 
preprint   arXiv:1406.0307.
\bibitem{levin} M.~Levin, Z.~C.~Gu 
  ``Braiding statistics approach to symmetry-protected topological phases'', 
  Phys. Rev. B {\bf 86} (2012) 115109; 
  M. Levin, 
  ``Protected edge modes without symmetry'', 
Phys. Rev. X {\bf 3} (2013) 021009; 
  Z.~C.~Gu, M.~Levin, 
  ``The effect of interactions on 2D fermionic symmetry-protected 
  topological phases with ${Z}_{2}$ symmetry'', 
Phys. Rev. B {\bf 89} (2014) 201113.
   \bibitem{vish2} M.A. Metlitski, L. Fidkowski, X. Chen, A. Vishwanath, 
``Interaction effects on 3D topological superconductors: 
surface topological order from vortex condensation, 
the 16 fold way and fermionic Kramers doublets'', 
preprint arXiv:1406.3032.
  \bibitem{kane} 
  C.~L.~Kane, E.~J.~Mele, 
  ``${Z}_{2}$ Topological Order and the Quantum Spin Hall Effect,'' 
  Phys.\ Rev.\ Lett.\  {\bf 95} (2005) 146802; 
  L.~Fu, C.~ L.~ Kane, 
  ``Time reversal polarization and a ${Z}_{2}$ adiabatic spin pump'', 
  Phys. Rev. B {\bf 74} (2006) 195312; 
  ``Topological insulators with inversion symmetry'', 
   Phys. Rev. B {\bf 76} (2007) 045302; 
  L.~Fu, C.~ L.~ Kane, E.~J.~Mele, 
  ``Topological Insulators in Three Dimensions'', 
  Phys. Rev. Lett. {\bf 98} (2007) 106803. 
\bibitem{tbt} 
  J. E. Moore, L. Balents, 
  ``Topological invariants of time-reversal-invariant band structures'', 
  Phys. Rev. B {\bf 75} (2007) 121306; 
  R. Roy, 
  ``$Z_2$ classification of quantum spin Hall systems: 
  An approach using time-reversal invariance'', 
  Phys. Rev. B {\bf 79} (2009) 195321. 
   \bibitem{qz} 
  X.~-L.~Qi, S.~-C.~Zhang, 
  ``Spin Charge Separation in the Quantum Spin Hall State,'' 
  Phys. Rev. Lett. {\bf 101} (2008) 086802. 
   \bibitem{laugh} 
  R.~B.~Laughlin, 
  ``Quantized Hall conductivity in two-dimensions,'' 
  Phys. Rev. B {\bf 23} (1981)  5632.
  \bibitem{cdtz}
  A.~Cappelli, G.~V.~Dunne, C.~A.~Trugenberger, G.~R.~Zemba, 
  ``Conformal symmetry and universal properties of quantum Hall states,'' 
  Nucl. Phys. B {\bf 398} (1993) 531. 
\bibitem{wen-book} 
  X. G. Wen, 
  Quantum Field Theory of Many-body Systems, 
  Oxford Univ. Press (2007), Oxford. 
   \bibitem{mr} 
  G.~W.~Moore, N.~Read, 
  ``Nonabelions in the fractional quantum hall effect,'' 
  Nucl. Phys. B {\bf 360} (1991) 362. 
\bibitem{rr} 
  N.~Read, E.~Rezayi, 
  ``Beyond paired quantum Hall states: Parafermions and 
  incompressible states in the first excited Landau level,'' 
  Phys. Rev. B {\bf 59} (1999) 8084.  
  \bibitem{nass} E. Ardonne, K. Schoutens, 
``New Class of Non-Abelian Spin-Singlet Quantum Hall States'', 
Phys. Rev. Lett. {\bf 82} (1999) 5096; 
E. Ardonne, N. Read, E. Rezayi, K. Schoutens, 
``Non-abelian spin-singlet quantum Hall states: wavefunctions 
and quasihole state counting'', 
Nucl. Phys. B {\bf 607} (2001) 549;
K. Schoutens, E. Ardonne, \& F.~J.~M. van Lankvelt, 
``Paired and Clustered Quantum Hall States'', 
in Proc. of the NATO Advanced Research Workshop "Statistical Field Theories" 
Como (Italy), June 18-23, 2001, A. Cappelli, G. Mussardo, eds. 
(Kluwer Academic Publishers 2002);  
E. Ardonne, K. Schoutens, 
``Wavefunctions for topological quantum registers'', 
Ann. Phys. {\bf 322}, (2007) 201.
  \bibitem{cz}
  A.~Cappelli and G.~R.~Zemba,
  ``Modular invariant partition functions in the quantum Hall effect,''
  Nucl.\ Phys.\ B {\bf 490} (1997) 595.
  \bibitem{cv} A.~Cappelli, G.~Viola, 
  ``Partition Functions of Non-Abelian Quantum Hall States,'' 
  J. Phys. A {\bf 44} (2011) 075401.
  \bibitem{cft} 
  P. Di Francesco, P. Mathieu, D. S\'en\'echal, 
  Conformal Field Theory, 
  Springer-Verlag (1997) New York.
    \bibitem{anom} 
  E. Witten, 
  ``An $SU(2)$ anomaly'' 
  Phys. Lett. B {\bf 117} (1982) 324; 
 A.~J.~Niemi, G.~W.~Semenoff,
 ``Axial Anomaly Induced Fermion Fractionization and Effective Gauge
Theory Actions in Odd Dimensional Space-Times,''
 Phys.\ Rev.\ Lett.\  {\bf 51}, 2077 (1983);
   A. N. Redlich, 
  ``Parity Violation and Gauge Noninvariance of the Effective 
  Gauge Field Action in Three-Dimensions'' 
  Phys. Rev. D {\bf 29} (1984) 2366; 
  S.~Deser, L.~Griguolo, D.~Seminara, 
  ``Effective QED actions: Representations, gauge invariance, 
  anomalies and mass expansions,'' 
  Phys. Rev. D {\bf 57} (1998) 7444. 
   \bibitem{rz}
  S.~Ryu, S.~C.~Zhang,
  ``Interacting topological phases and modular invariance,''
  Phys. Rev. B {\bf 85} (2012) 245132.
   \bibitem{cgt1} 
  A.~Cappelli, L.~S.~Georgiev, I.~T.~Todorov, 
  ``A Unified conformal field theory description of paired quantum Hall 
  states,'' 
  Commun. Math. Phys. {\bf 205} (1999) 657. 
\bibitem{cgt2} 
  A.~Cappelli, L.~S.~Georgiev, I.~T.~Todorov, 
  ``Parafermion Hall states from coset projections of Abelian 
  conformal theories,'' 
  Nucl. Phys. B {\bf 599} (2001) 499. 
   \bibitem{hal} 
F. D. M. Haldane, 
``Stability of Chiral Luttinger Liquids and Abelian Quantum Hall States",
Phys. Rev. Lett. {\bf 74} (1995) 2090.
\bibitem{zamo}
Al. B. Zamolodchikov, 
``From tricritical Ising to critical Ising by thermodynamic Bethe ansatz'',
Nucl. Phys. B {\bf 358} (1991) 524; 
 A. B. Zamolodchikov, 
``Expectation value of composite field $T{\bar T}$ in two-dimensional 
quantum field theory", preprint arXiv:hep-th/0401146;
M. Caselle, D. Fioravanti, F. Gliozzi, R. Tateo, 
``Quantisation of the effective string with TBA", preprint arXiv:1305.1278.
\bibitem{wen-na} M. Barkeshli, X.G. Wen, 
``Effective field theory and projective construction for 
$Z_k$ parafermion fractional quantum Hall states'', 
Phys. Rev. B {\bf 81} (2010) 155302.
\end{thebibliography}
\end{document}